\documentclass[review]{elsarticle}

\usepackage{graphicx}
\usepackage{dcolumn}
\usepackage{bm}
\usepackage{epsfig}
\usepackage{booktabs}
\usepackage{subfigure}
\usepackage{graphics}
\usepackage{amssymb}
\usepackage{amsmath}
\usepackage{array}
\usepackage{color}

\journal{Computers \& Mathematics with Applications}

\begin{document}

\begin{frontmatter}

\title{A generalized lattice Boltzmann model for fluid flow system and its application in two-phase flows}
\author[mymainaddress]{Xiaolei Yuan}
\author[mymainaddress,mysecondaryaddress]{Zhenhua Chai}
\author[mythirdaryaddress]{Huili Wang}
\author[mymainaddress,mysecondaryaddress]{Baochang Shi\corref{mycorrespondingauthor}}
\cortext[mycorrespondingauthor]{Corresponding author}
\ead{shibc@hust.edu.cn}

\address[mymainaddress]{School of Mathematics and Statistics, Huazhong University of Science and Technology, Wuhan 430074, China}
\address[mysecondaryaddress]{Hubei Key Laboratory of Engineering Modeling and Scientific Computing, Huazhong University of Science and Technology, Wuhan
    430074, China}
\address[mythirdaryaddress]{School of Mathematics and Computer Science, Wuhan Textile University, Wuhan, 430073, China}

\begin{abstract}
In this paper, a generalized lattice Boltzmann (LB) model with a
mass source is proposed to solve both incompressible and nearly
incompressible Navier-Stokes (N-S) equations. This model can be used
to deal with single-phase and two-phase flows problems with a mass
source term. From this generalized model, we can not only get some
existing models, but also derive new models. Moreover, for the
incompressible model derived, a modified pressure scheme is
introduced to calculate the pressure, and then to ensure the
accuracy of the model. In this work, we will focus on a two-phase
flow system, and in the frame work of our generalized LB model, a
new phase-field-based LB model is developed for incompressible and
quasi-incompressible two-phase flows. A series of numerical
simulations of some classic physical problems, including a spinodal
decomposition, a static droplet, a layered Poiseuille flow, and a
bubble rising flow under buoyancy, are performed to validate the
developed model. Besides, some comparisons with previous
quasi-incompressible and incompressible LB models are also carried
out, and the results show that the present model is accurate in the
study of two-phase flows. Finally, we also conduct a comparison
between quasi-incompressible and incompressible LB models for
two-phase flow problems, and find that in some cases, the proposed
quasi-incompressible LB model performs better than incompressible LB
models.
\end{abstract}


\end{frontmatter}

\section{\label{sec:level1}Introduction}

Fluid flow problems are ubiquitous in nature and engineering
applications, such as single and multiphase flows
\cite{Tryggvason,He0,Aursjo0,Ramstad,Jiang,Liang1}, thermal flows
\cite{Vafai,Peng}, and porous media flows \cite{Pan,Jiang}. For such
problems, the phenomenon of fluid flowing in and out of the system
is very frequent. In order to describe this phenomenon, an
alternative way is to introduce a mass source or sink term in the
fluid dynamical equations (i.e., N-S equations)
\cite{Aursjo,Kuzmin,Crowe,Migdal}. Besides, in the presence of
electro-chemical reactions or multiphase situations, mass sources in
the N-S equations could be applied to describe the reaction between
the components in the system \cite{Aursjo,Dutta}. Due to limitations
of theoretical research and experimental methods, it is especially
necessary to develop a numerical method to solve the N-S equations
with a mass source term.

As an efficient numerical method, the lattice Boltzmann method has
made rapid progress since its appearance in the late 1980s due to
its simplicity, scalability on parallel computers, and ease to
handle complex geometries \cite{Kruger,Chai0}. This method has
gained great success in the study of fluid flow system
\cite{He1,Guo1,Aidun,Liang2019}. In general, LB methods for dealing
with such flow problems can be divided into two categories. One is
the nearly incompressible model and the other is the so-called
incompressible model. For the nearly incompressible model, the
macroscopic quantities those need to be solved are the density
$\rho$ and fluid velocity $\mathbf{u}$, and the pressure can be
obtained from the equation of state ($p=\rho c_s^2$). While for the
incompressible model, the macroscopic quantities we need to
calculate are fluid velocity $\mathbf{u}$ and the pressure $p$,
where density $\rho$ is viewed as a constant. In the past, people
always considered these two types of models separately. Actually,
from the perspective of model construction, one can design a
generalized LB model to deal with both incompressible and nearly
incompressible problems, which is the main motivation of this paper.
Moreover, some previous models have more or less assumptions in the
derivation process, and often cannot recover the macroscopic
equations completely. In addition, under the framework of LB
methods, people have less research on N-S equations with a mass source term.
For instance, Halliday \emph{et al.} \cite{Halliday} proposed a
single-relaxation-time (SRT) LB model including a mass source term,
while they employed a non-local scheme to calculate the spatial
derivatives which appear in the source term. Cheng \emph{et al.}
\cite{Cheng1} presented another LB model with  a general mass source
term and adopted a non-local scheme for temporal and spatial
derivatives. Aursj\o \ \emph{et al.} \cite{Aursjo} also developed an
SRT model with a mass source term, which does not include temporal
and spatial derivatives in the source term, and it preserves the
Galilean invariance. We note that the above models are limited to
nearly incompressible situations. While, up to now, there is no
available work on incompressible N-S equations with a mass source
term. 
Considering the above points, in this work, we will develop a
generalized SRT LB model for both incompressible and nearly
incompressible N-S equations with a mass source term, and this model
is also an extension of existing models. From the generalized model,
we can not only get some existing models, but also derive new
models. Among these new models, we can obtain an incompressible
model for N-S equations with a mass source term, and we present a
modified scheme to calculate the pressure $p$, which is more
accurate than the previous one. Simultaneously, our generalized
model can recover
the macroscopic equations without any unnecessary assumptions. 
Finally, we would like to point that this
 model can be used not only to solve single-phase
fluid flows, but also to design models for two-phase flows.

Based on our generalized model, we also present a phase-field-based
LB model for two-phase flows. The present model contains both
quasi-incompressible and incompressible situations, in which the
quasi-incompressible model can guarantee the mass conservation, and
its governing equation of the flow field can be regarded as a kind
of incompressible N-S equation with a mass source term. Actually,
there are also some phase-field-based LB models for incompressible
two-phase flows. He \emph{et al.} \cite{He0} proposed a phase-field
LB model and adopted an order parameter to track the interface of
two incompressible fluids. However, there are some differences
between the derived governing equations and the phase-field theory
for incompressible two-phase flows \cite{Zu}. To recover the
Cahn-Hilliard (CH) equation correctly, Zheng \emph{et al.}
\cite{Zheng1} and Zu \emph{et al.} \cite{Zu} developed two different
LB models, while, the extra terms in the recovered macroscopic
equations from their models will produce large errors in the
interface capturing, and numerical instability will occur when the
dimensionless relaxation time equals to $1$ \cite{Zu}. To overcome
these problems, Liang \emph{et al.} \cite{Liang} proposed a new
multi-relaxation-time (MRT) LB model through introducing a
time-dependent source term in the evolution equation of phase field.
While all the above models cannot conserve mass locally when the
densities of the two fluids are different. To solve the problem,
Yang \emph{et al.} \cite{Yang} presented a modified LB model based
on the quasi-incompressible theory. From his model, the
quasi-incompressible N-S equations in artificial compressible form
can be derived. To neglect the artificial compressible effect, the
model requires an additional condition, $T \gg L/c_s$ ($T$ and $L$
are characteristic time and length, respectively). Based on the work
of Yang \emph{et al.} \cite{Yang}, Zhang \emph{et al.}
\cite{Zhangchunhua} proposed a discrete unified gas-kinetic scheme
(DUGKS) for two-phase flows which can exactly guarantee the mass
conservation, while this model also give rise to the generation of
artificial compressible effect. On the contrary, the present
phase-field-based LB model can overcome the above defects. The rest
of present paper is organized as follows. In Sec.
\uppercase\expandafter{\romannumeral2}, the generalized LB model for
fluid flow system with a mass source is introduced, and a
phase-field-based LB model for two-phase flows is given in Sec.
\uppercase\expandafter{\romannumeral3}. Numerical experiments to
validate the present model are carried out in Sec.
\uppercase\expandafter{\romannumeral4}. Finally, some conclusions
are given in Sec. \uppercase\expandafter{\romannumeral5}.

\section{Generalized LB model for fluid flow system with a mass
source}
\subsection{Governing equations}

The governing equations for nearly incompressible fluid flows with a
mass source term 
 can be written as \cite{Aursjo,Unverdi,Liang2018}
\begin{subequations}
\begin{equation}
\frac{\partial {\rho}}{\partial t}+ \nabla \cdot \left(\rho
\textbf{u}\right) = S,
 \label{eq:001}
\end{equation}
\begin{equation}
\begin{split}
\frac{\partial (\rho \textbf{u})}{\partial t} + \nabla \cdot
\left(\rho \textbf{u} \textbf{u}\right) = & - \nabla \emph{p}+\nabla
\cdot \tau+\textbf{F},
\end{split}
\label{eq:002}
\end{equation}
\label{eq:002a}
\end{subequations}
where $\rho$ is the density, $\textbf{u}$ denotes the fluid
velocity, $S$ is a mass source or sink term, $\emph{p}$ is the
hydrodynamic pressure, $\textbf{F}$ is the external force, $\tau$ is
the deviatoric stress tensor, and for Newtonian fluids,
\begin{equation}
\begin{split}
\tau = &  \mu (\nabla \textbf{u}+\nabla \textbf{u}^\emph{T})
+\left(\xi-\frac{2}{d} \mu\right) \nabla \cdot \mathbf{u} \mathbf{I}
\end{split}
\label{eq:3}
\end{equation}
where $\mu$ denotes the dynamic viscosity by $\mu=\rho \nu$, $\nu$
is the kinematic viscosity, $\xi$ is the bulk (or volume) viscosity
and $d$ is the number of spatial dimensions.

If we take $S=0$ in Eq. (\ref{eq:002a}), the governing equations for
nearly incompressible fluid flow system can be obtained
\cite{Kruger},
\begin{subequations}
\begin{equation}
\frac{\partial {\rho}}{\partial t}+ \nabla \cdot \left(\rho
\textbf{u}\right) = 0,
 \label{eq:01}
\end{equation}
\begin{equation}
\begin{split}
\frac{\partial (\rho \textbf{u})}{\partial t} + \nabla \cdot
\left(\rho \textbf{u} \textbf{u}\right) = & - \nabla \emph{p}+\nabla
\cdot \tau+\textbf{F},
\end{split}
\label{eq:02}
\end{equation}
\label{eq:02a}
\end{subequations}

For fluid flows with small temperature changes, the flow can be
regarded as incompressible under the condition of $\rho=const$, so
that the above governing equations will reduce to \cite{Kruger,Guo1}
\begin{subequations}
\begin{equation}
\nabla \cdot \textbf{u} =0,
\end{equation}
\begin{equation}
\begin{split}
\frac{\partial  \textbf{u}}{\partial t} + \nabla \cdot \left(
\textbf{u} \textbf{u}\right) = & - \nabla \emph{p}+\nabla \cdot
[\nu(\nabla \mathbf{u})]+\mathbf{F}.
\end{split}
\end{equation}
\label{eq:03a}
\end{subequations}

In this work, to simplify the following discussion and facilitate
the design of a generalized model, here we consider the following
generalized governing equations with a mass source,
\begin{subequations}
\begin{equation}
\frac{\partial \tilde{\rho}}{\partial t}+ \nabla \cdot \left({\rho}
\textbf{u}\right) = S,
 \label{eq:1}
\end{equation}
\begin{equation}
\begin{split}
\frac{\partial ({\rho} \textbf{u})}{\partial t} + \nabla \cdot
\left({\rho} \textbf{u} \textbf{u}\right) = & - \nabla
\emph{p}+\nabla \cdot \tau+\textbf{F},
\end{split}
\label{eq:2}
\end{equation}
\label{eq:2a}
\end{subequations}
 where $\tilde{\rho}$ is physical quantity related to $\rho$ or a constant.
  Note that Eqs. (\ref{eq:2a}) are a kind of generalized N-S equations
  which contain two basic forms of incompressible and nearly incompressible
 governing equations, and they can be used to describe single-phase
or multi-phase flows with a mass source term. When $\tilde{\rho}$
and $S$ take different values, Eqs. (\ref{eq:2a}) represents
different governing equations.  For example, if we take
$\tilde{\rho}=\rho$, Eqs. (\ref{eq:2a}) will reduce to Eqs.
(\ref{eq:002a}), and further reduce to Eqs. (\ref{eq:02a}) with
$S=0$. While if we take $\tilde{\rho}=\rho=const$,  and $S=0$, Eqs.
(\ref{eq:03a}) can be derived. Next, we will design the
corresponding LB model for Eqs (\ref{eq:2a}), and the designed model
must be able to deal with the incompressible and near incompressible
single-phase and multi-phase flow problems with a mass source term.
From this perspective, the LB model we designed is also a
generalized model.

\subsection{LB model for generalized N-S equations with a mass source term}

To obtain the evolution equation of Eq. (\ref{eq:2a}), we integrate
the following discrete velocity Boltzmann equation
\begin{equation}
\frac{\partial f_i}{\partial t}+\mathbf{c}_i \cdot \nabla
f_i=\Omega_i+G_i \label{eq:20}
\end{equation}
along a characteristic line $\mathbf{c}_i$ over a time interval
$\Delta t$ \cite{Luo,Du}, and we have
\begin{equation}
f_i(\textbf{x}+\textbf{c}_i \Delta t,t+\Delta t)-f_i(\textbf{x},t) =
\int_0^{\Delta t} \Omega_i(\textbf{x}+\textbf{c}_i t',t+t') d
t'+\int_0^{\Delta t} G_i(\textbf{x}+\textbf{c}_i  t',t+ t') d t'.
\label{eq:21}
\end{equation}
where $f_i(\mathbf{x},t)$ denotes particle distribution function
with velocity $\mathbf{c}_i$ at position $\mathbf{x}$ and time $t$,
$G_i(\mathbf{x},t)$ represents the force term,
$\Omega_i(\mathbf{x},t)$ is the collision operator approximated by
\begin{equation}
\Omega_i=-\frac{1}{\tau'}(f_i-f_i^{eq}),
\end{equation}
where $\tau'$ is the relaxation time and $f_i^{eq}(\mathbf{x},t)$ is
the equilibrium distribution function.

The integral of the collision term adopts the trapezoidal rule, then
Eq. (\ref{eq:21}) becomes
\begin{equation}
f_i(\textbf{x}+\textbf{c}_i \Delta t,t+\Delta t)-f_i(\textbf{x},t)
=\frac{\Delta t}{2} \left[\Omega_i(\textbf{x}+\textbf{c}_i \Delta
t,t+\Delta t)+\Omega_i(\textbf{x},t)\right] +\int_0^{\Delta t}
G_i(\textbf{x}+\textbf{c}_i t',t+t') d t'. \label{eq:21b}
\end{equation}

Let $\bar{f}_i=f_i-\frac{\Delta t}{2}\Omega_i$, we have
\begin{equation}
\bar{f}_i(\textbf{x}+\textbf{c}_i \Delta t,t+\Delta
t)-\bar{f}_i(\textbf{x},t) =-\frac{1}{\tau_g}\left[
\bar{f}_i(\mathbf{x},t)-f_i^{eq}(\mathbf{x},t)\right]
+\int_0^{\Delta t} G_i(\textbf{x}+\textbf{c}_i t',t+t') d t',
\label{eq:21bb}
\end{equation}
where $\tau_g=\frac{2\tau'+\Delta t}{2\Delta t}$ is the
dimensionless relaxation time, and $\bar f_i$ satisfies $\sum_i \bar
f_i=\sum_i f_i$, and $\sum_i \mathbf{c}_i \bar f_i=\sum_i
\mathbf{c}_i f_i$.

Through Taylor expansion of $G_i$ and neglecting the terms of order
$O(\Delta t^2)$, the last term in the right hand of Eq.
(\ref{eq:21bb}) becomes
\begin{equation}
\begin{split}
\int_0^{\Delta t} G_i(\textbf{x}+\textbf{c}_i \Delta t',t+\Delta t')
d t' &=\int_0^{\Delta t} \left(G_i(\mathbf{x},t)+t' D_i
G_i(\mathbf{x},t)\right)dt' \\
&=\Delta t G_i(\mathbf{x},t)+\frac{\Delta t^2}{2}D_i
G_i(\mathbf{x},t),
\end{split}
\label{eq:22}
\end{equation}
where $D_i=\partial t+\mathbf{c}_i \cdot \nabla$.

The LB evolution equation with the BGK collision operator for the
N-S equations can be expressed as \cite{Du}
\begin{equation}
\bar{f}_i(\textbf{x}+\textbf{c}_i \Delta t,t+\Delta
t)-\bar{f}_i(\textbf{x},t) =-\frac{1}{\tau_g}\left[
\bar{f}_i(\mathbf{x},t)-f_i^{eq}(\mathbf{x},t)\right] +\Delta t
\left[G_i(\mathbf{x},t)+\frac{\Delta t}{2}D_i G_i \right],
\label{eq:23}
\end{equation}
If the up-wind scheme is used to Eq. (\ref{eq:23}), the evolution
equation for the generalized N-S equations reads
\begin{equation}
\bar{f}_i(\textbf{x}+\textbf{c}_i \Delta t,t+\Delta
t)-\bar{f}_i(\textbf{x},t) =-\frac{1}{\tau_g}\left[
\bar{f}_i(\mathbf{x},t)-f_i^{eq}(\mathbf{x},t)\right]+\Delta t
\left[G_i(\mathbf{x},t)+\frac{G_i(\textbf{x}+\textbf{c}_i \Delta
t,t+\Delta t)-G_i(\textbf{x},t)}{2} \right].
\end{equation}

To remove the implicitness, we introduce a modified particle
distribution function,
\begin{equation}
g_i(\mathbf{x},t)=\bar f_i(\mathbf{x},t)-\frac{\Delta t}{2}G_i.
\label{eq:23a}
\end{equation}
With some simple manipulations, the explicit evolution equation can
be derived,
\begin{equation}
g_i(\textbf{x}+\textbf{c}_i \Delta t,t+\Delta t)-g_i(\textbf{x},t)
=-\frac{1}{\tau_g}
\left(g_i(\mathbf{x},t)-g_i^{eq}(\mathbf{x},t)\right)+\Delta
t(1-\frac{1}{2\tau_g})G_i, \label{eq:24}
\end{equation}
where $g_i^{eq}$ is the new equilibrium distribution function and
satisfies $g_i^{eq}=f_i^{eq}$.

To derive Eq. (\ref{eq:2a}) through Chapman-Enskog analysis, the
equilibrium distribution function $g_i^{eq}$ is defined as (see
\ref{app:sec1} for the details)
\begin{equation}
g_i^{eq}=\left\{
\begin{array}{l}
\tilde{\rho}+\frac{p}{c_s^2}(\omega_i-1)+{\rho} s_i(\mathbf{u}), \;\;\;i=0\\
\frac{p}{c_s^2}\omega_i+{\rho} s_i(\mathbf{u}),
\;\;\;\;\;\;\;\;\;\;\;\;\;\;\;\;\;   i \ne 0,
\end{array}
\right. \label{eq:25}
\end{equation}
with
\begin{equation}
s_i({\mathbf{u}})=\omega_i \left[ \frac{\mathbf{c}_i \cdot
\mathbf{u}}{c_s^2}+\frac{(\mathbf{c}_i \cdot
\mathbf{u})^2}{2c_s^4}-\frac{\mathbf{u} \cdot \mathbf{u}}{2c_s^2}
\right],
\end{equation}
where $\omega_i$ denotes the weighting coefficient, $\mathbf{c}_i$
is the discrete velocity, and $c_s$ is the speed of sound. Note that
our model is based on the D$d$Q$q$ lattice with $q$ velocity
directions in $d$-dimensional space. $\mathbf{c}_i$ and $\omega_i$
depend on the choice of the lattice model, and in D$1$Q$3$ model,
$\{\mathbf{c}_i\}=(0,1,-1)c$, $\omega_0=2/3$, $\omega_{1,2}=1/6$,
$c_s=c/\sqrt{3}$, where $c=\Delta x / \Delta t$ , with $\Delta x$
and $\Delta t$ representing the spacing and time step, respectively;
while in the D$2$Q$9$ model, $\omega_i$ is given by $\omega_0=4/9$,
$\omega_{1-4}=1/9$, $\omega_{5-8}=1/36$, $c_s=c/\sqrt{3}$, and
$\mathbf{c}_i$ is given by
$$\{\mathbf{c}_i\}=\left( \begin{array}{rrrrrrrrr}
0&1&0&-1&0&1&-1&-1&1\\
0&0&1&0&-1&1&1&-1&-1 \end{array} \right)c;$$ in the D$3$Q$15$ model,
 $\omega_0=2/9$, $\omega_{1-6}=1/9$,
$\omega_{7-14}=1/72$, $c_s=c/\sqrt{3}$, and $\mathbf{c}_i$ is given
by
$$\{\mathbf{c}_i\}=\left( \begin{array}{rrrrrrrrrrrrrrr}
0&1&0&0&-1&0&0&1&1&1&-1&-1&-1&-1&1\\
0&0&1&0&0&-1&0&1&1&-1&1&-1&-1&1&-1\\
0&0&0&1&0&0&-1&1&-1&1&1&-1&1&-1&-1 \end{array} \right)c.$$

Different from some previous LB models
\cite{He2,Yang,Liang,Lee1,Lee2,Zu,Ren,Fakhari2017}, in the present
model, the force distribution function is given by
\begin{equation}
G_i=\omega_i\left\{ S+\frac{\mathbf{c}_i \cdot
\mathbf{F}}{c_s^2}+\frac{(\mathbf{c}_i
\mathbf{c}_i-c_s^2\mathbf{I}):\left[\partial_t(p-\tilde{\rho}c_s^2)\mathbf{I}+\mathbf{u
\tilde{F}}+\mathbf{ \tilde{F}u}-\mathbf{uu}\tilde{S}+Q(\nabla \cdot
\mathbf{uI})\right]}{2c_s^4}\right \} , \label{eq:27}
\end{equation}
where $\tilde{\mathbf{F}}$ is a modified total force
\begin{equation}
\mathbf{\tilde{F}}=\mathbf{F}-\nabla(p-{\rho} c_s^2),
\end{equation}
$\tilde{S}$ and $Q$ can be expressed as
\begin{equation}
\tilde{S}=S+\partial_t({\rho}-\tilde{\rho}).
\end{equation}
\begin{equation}
Q=\frac{2}{d}{\rho} c_s^2-\frac{\xi}{\Delta t(\tau_g-0.5)}.
\end{equation}
Under the Stokes hypothesis, the bulk viscosity $\xi$ is usually
assumed to be zero \cite{George}. In addition, $\xi$ can also take
$2\rho\nu/d$, so that the derived macroscopic equations will not
contain the term $\left(\xi-\frac{2}{d} \mu\right) \nabla \cdot
\mathbf{u} \mathbf{I}$. Please note that $G_i$ is a complete form
and doesn't contain any unnecessary assumptions, and $G_i$ can be
reasonably simplified for specific problems.

We would like to point out that through the Chapman-Enskog analysis,
the present LB model with the force term Eq. (\ref{eq:27}) can
correctly recover Eq. (\ref{eq:2a}) (see \ref{app:sec1} for the
details), and simultaneously, the fluid kinematic viscosity can be
determined by
\begin{equation}
\nu=c_s^2(\tau_g-0.5)\Delta t.
\end{equation}

In the implementation of the present model, for nearly
incompressible model, the macroscopic quantities $\tilde{\rho}$ can
be calculated as
\begin{equation}
\tilde{\rho}=\sum_i g_i+\frac{\Delta t}{2}S, \label{eq:27aa}
\end{equation}
where Eq. (\ref{eq:23a}) is used. By taking the first-order moment
of $g_i$, the fluid velocity can be obtained
\cite{Liang,Liang1,Liang2018},

\begin{equation}
\mathbf{u}=\frac{1}{{\rho}} \left( \sum_i \mathbf{c}_i g_i+0.5\Delta
t \mathbf{F} \right). \label{eq:27a}
\end{equation}
 While for the incompressible model,
the macroscopic quantities we need to calculate are fluid velocity
$\mathbf{u}$ and the pressure $p$. The fluid velocity is given by
Eq. (\ref{eq:27a}), and the pressure can be calculated as (see
\ref{app:sec2} for the details)
\begin{equation}
p=\frac{c_s^2}{1-\omega_0} \left[ \sum_{i\ne 0} g_i +\frac{\Delta
t}{2}S+{\rho} s_0(\mathbf{u})+\Delta t(\tau_g-\frac{1}{2})G_0 \right
], \label{eq:28}
\end{equation}
or
\begin{equation}
p=\frac{c_s^2}{1-\omega_0} \left[ \sum_{i\ne 0} g_i +\frac{\Delta
t}{2}S+{\rho} s_0(\mathbf{u})+\Delta
t(\tau_g-\frac{1}{2})(S-\sum_{i\ne 0}G_i) \right ]. \label{eq:28aa}
\end{equation}

We would also like to point out that the present pressure expression
is different from that in Refs. \cite{Liang,Liang2018,Ren}  which
can be expressed as
\begin{equation}
p=\frac{c_s^2}{1-\omega_0} \left[ \sum_{i\ne 0} g_i +\frac{\Delta
t}{2}S+\rho s_0(\mathbf{u}) \right ], \label{eq:29}
\end{equation}
where the term of $\Delta t(\tau_g-\frac{1}{2})G_0$ has been
neglected. However, the last term $\Delta t(\tau-\frac{1}{2})G_0$
may be significant since  $G_0$ is usually nonzero under the
condition of $S \ne 0$, and the effect of this item cannot be
ignored.

Thus, our generalized LB model for fluid flow system are made up of
Eqs. (\ref{eq:24}), (\ref{eq:25}), (\ref{eq:27}) and
(\ref{eq:27aa}), (\ref{eq:27a}) or (\ref{eq:27a}), (\ref{eq:28}). In
the derivation of the model, we only used the assumption of low Mach
number ($Ma\ll1$), which is necessary for the construction of most
LB models. Under this assumption, the following equations are
established for nearly incompressible or incompressible flows,
$\mathbf{u}=O(Ma)$, $\delta \rho=O(Ma^2)$, and $\delta p=O(Ma^2)$.
Further, we have $\mathbf{F}=O(Ma)$ and $S=O(Ma)$. Now, we will use
some remarks to show that our model not only contains some existing
models, but also deduces new models for N-S equations with a mass
source term.

\emph{Remark 1.} When taking $\tilde{\rho}=\rho$ and $p=\rho c_s^2$,
the generalized model can reduce to the nearly incompressible form.
Correspondingly, the macroscopic equations becomes Eqs.
(\ref{eq:002a}), while the equilibrium distribution function can be
written as
\begin{equation}
g_i^{eq}=\rho\left[ \omega_i+s_i(\mathbf{u})\right], \label{eq:29bb}
\end{equation}
and the force distribution function is given by
\begin{equation}
G_i=\omega_i\left\{ S+\frac{\mathbf{c}_i \cdot
\mathbf{F}}{c_s^2}+\frac{(\mathbf{c}_i
\mathbf{c}_i-c_s^2\mathbf{I}):\left[\mathbf{u {F}}+\mathbf{
{F}u}-\mathbf{uu}{S}+Q (\nabla \cdot
\mathbf{uI})\right]}{2c_s^4}\right \} ,
\end{equation}
where the term $\mathbf{uu}{S}$ is the order of $O(Ma^3)$ and can be
neglected. This model is almost identical to the one in Ref.
\cite{Aursjo}, except that the expression of the momentum equation
and the value of bulk viscosity are different.

\emph{Remark 2.} If we rewrite the equilibrium distribution function
Eq. (\ref{eq:29bb}) to the following form,
\begin{equation}
g_i^{eq}=\omega_i \left\{\rho+(\rho_0+\delta \rho)\left[
\frac{\mathbf{c}_i \cdot \mathbf{u}}{c_s^2}+\frac{(\mathbf{c}_i
\cdot \mathbf{u})^2}{2c_s^4}-\frac{\mathbf{u} \cdot
\mathbf{u}}{2c_s^2}\right] \right \}. \label{eq:29bc}
\end{equation}
Multiplying by $c_s^2$ on both sides of Eq. (\ref{eq:29bc}) and
ignoring the term of $O(Ma^2)$, one can get
\begin{equation}
g_i^{eq}=\omega_i \left\{p+p_0\left[ \frac{\mathbf{c}_i \cdot
\mathbf{u}}{c_s^2}+\frac{(\mathbf{c}_i \cdot
\mathbf{u})^2}{2c_s^4}-\frac{\mathbf{u} \cdot
\mathbf{u}}{2c_s^2}\right] \right \}, \label{eq:29bd}
\end{equation}
where we still use $g_i^{eq}$ to represent the new equilibrium
distribution function. The force distribution function is given by
\begin{equation}
G_i=\omega_i\left\{ p_0 S_1+\frac{p_0 \mathbf{c}_i \cdot
\mathbf{F_1}}{c_s^2}+\frac{(\mathbf{c}_i
\mathbf{c}_i-c_s^2\mathbf{I}):\left[p_0(\mathbf{u {F_1}}+\mathbf{
{F_1}u})+\left(\frac{2}{d}{p_0} c_s^2-\frac{\xi}{\Delta
t(\tau_g-0.5)} \right) (\nabla \cdot
\mathbf{uI})\right]}{2c_s^4}\right \} ,
\end{equation}
where $p_0=const$, $S_1=\frac{S}{p_0}$, $F_1=\frac{F}{p_0}$, and the
terms of $O(Ma^3)$ are abandoned in the incompressible limit. The
fluid velocity and pressure can be obtained by $p=\sum_i
g_i+\frac{\Delta t}{2}p_0 S_1$ and $p_0\mathbf{u}=\sum_i c_i
g_i+\frac{\Delta t}{2}p_0 \mathbf{F_1}$. The corresponding
incompressible N-S equations with a source term $S_1$ can be
obtained from Eqs. (\ref{eq:2a}),
\begin{subequations}
\begin{equation}
\frac{1}{c_s^2}\frac{\partial P}{\partial t}+\nabla \cdot \textbf{u}
= S_1,
 \label{eq:50}
\end{equation}
\begin{equation}
\begin{split}
\frac{\partial  \textbf{u}}{\partial t} + \nabla \cdot \left(
\textbf{u} \textbf{u}\right) = & - \nabla {P}+\nabla \cdot
[\nu(\nabla \mathbf{u}+\nabla \mathbf{u}^\emph{T})]+\nabla \cdot
[(-\frac{2}{3}\nu)\nabla \cdot \mathbf{uI}]+\textbf{F}_1,
\end{split}
\end{equation}
\end{subequations}
where  $P=\frac{p}{\rho_0}$. In the limit of a low Mach number
($Ma=|\mathbf{u}|/c_s^2$), the dynamic pressure is assumed to be
$\delta p \sim Ma^2$ and the left end term of Eq. (\ref{eq:50}) can
be ignored. Note that if we take $S_1=0$, this model can reduce to
the incompressible LB model by He \emph{et al.} \cite{He1}.

\emph{Remark 3.} 
 When taking $\tilde{\rho}=\rho_0$, ${\rho} = const$ (e.g., ${\rho}=1$), where $\rho_0$ is the
average velocity of the fluid, we can derive a new model for
incompressible N-S equations with a mass source term. In this case,
the macroscopic equations can be written as  [see Eqs.
(\ref{eq:2a})]
\begin{subequations}
\begin{equation}
\nabla \cdot \textbf{u} = S,
\end{equation}
\begin{equation}
\begin{split}
\frac{\partial  \textbf{u}}{\partial t} + \nabla \cdot \left(
\textbf{u} \textbf{u}\right) = & - \nabla \emph{p}+\nabla \cdot
[\nu(\nabla \mathbf{u}+\nabla \mathbf{u}^\emph{T})]+\nabla \cdot
[(-\frac{2}{3}\nu)\nabla \cdot \mathbf{uI}]+\textbf{F}.
\end{split}
\end{equation}
\end{subequations}
The corresponding equilibrium distribution function can be derived
from Eq. (\ref{eq:25})
\begin{equation}
g_i^{eq}=\left\{
\begin{array}{l}
\rho_0+\frac{p}{c_s^2}(\omega_i-1)+ s_i(\mathbf{u}), \;\;\;i=0\\
\frac{p}{c_s^2}\omega_i+ s_i(\mathbf{u}),
\;\;\;\;\;\;\;\;\;\;\;\;\;\;\;\;\;\;   i \ne 0,
\end{array}
\right.
\end{equation}
and the force distribution function is given by
\begin{equation}
G_i=\omega_i\left\{ S+\frac{\mathbf{c}_i \cdot
\mathbf{F}}{c_s^2}+\frac{(\mathbf{c}_i
\mathbf{c}_i-c_s^2\mathbf{I}):\left[\mathbf{u {F}}+\mathbf{ {F}u}+Q
S \mathbf{I}\right]}{2c_s^4}\right \} ,
\end{equation}
where the term $\mathbf{uu}{\tilde{S}}$ and $\partial_t p$ are
neglect in the incompressible limit.

 We note that if we take $S=0$, this model can
reduce to the incompressible LB model by Guo \emph{et al.}
\cite{Guo1} with a force term, i.e., Eqs. (\ref{eq:03a}).

\emph{Remark 4.} Eq. ({\ref{eq:27a}}) may be implicit if the force
$\mathbf{F}$ is a nonlinear function of $\mathbf{u}$. To remove this
implicitness, we can discretize $D_i G_i$ by \cite{Du}
\begin{equation}
D_i G_i=\frac{G_i(\textbf{x}+\textbf{c}_i \Delta
t,t)-G_i(\textbf{x},t-\Delta t)}{\Delta t}.
\end{equation}
Then the evolution equation can be rewritten as
\begin{equation}
g_i(\textbf{x}+\textbf{c}_i \Delta t,t+\Delta t)-g_i(\textbf{x},t)
=-\frac{1}{\tau_g}
\left(g_i(\mathbf{x},t)-g_i^{eq}(\mathbf{x},t)\right)+\Delta t
\left[ G_i(\mathbf{x},t)+\frac{G_i(\textbf{x}+\textbf{c}_i \Delta
t,t)-G_i(\textbf{x},t-\Delta t)}{2} \right].
\end{equation}
This scheme is explicit and can be iterated if $G_i$ is known, while
the results of $G_i$ at time $t-\Delta t$ must be saved
additionally.

\section{Phase-field-based LB model for two-phase flows}

In Sec. \uppercase\expandafter{\romannumeral2}, we have given the
generalized LB model for single-phase flows with a mass source term.
Actually, one of the motivations of the generalized model is to deal
with the two-phase flow problems based on phase-field theory.
\subsection{Governing equations}
In the classic theory of phase field model for two-phase flows, the
N-S equations are employed to describe the flow field, while the
Cahn-Hilliard (C-H) equation is usually adopted to capture the phase
interface which can be given by \cite{Jacqmin,Kendon,Badalassi}
\begin{equation}
\frac{\partial \phi}{\partial t}+\nabla \cdot \phi \textbf{u}=\nabla
\cdot M_{\phi} \left( \nabla \mu \right),
 \label{eq:3a}
\end{equation}
where $M_{\phi}$ is the mobility coefficient, $\phi$ is the order
parameter defined as the volume fraction of one of the two phases,
and $\phi$ and $\rho$ satisfy the linear relationship,
\begin{equation}
\frac{\rho-\rho_B}{\rho_A-\rho_B}=\frac{\phi-\phi_B}{\phi_A-\phi_B}.
\label{eq:3aa}
\end{equation}
In this work, $\phi = 0$ denotes the phase B while $\phi = 1$
represents the phase A, and $\mu$ is the chemical potential, which
is derived by the variation of the free-energy function $F(\phi)$
with respect to
 the order parameter \cite{Jacqmin,Jacqmin1,Yan},
\begin{equation}
\begin{split}
\mu & =\frac{\delta F(\phi)}{\delta
\phi}=\frac{d\psi(\phi)}{d\phi}-\kappa
\nabla^2 \phi \\
&=4\beta\phi(\phi-1)(\phi-0.5)-\kappa \nabla^2 \phi.
\end{split}
\label{eq:8}
\end{equation}
where $F(\phi)$ 
can be taken as the following form \cite
{Jacqmin,Kendon,Badalassi,Chai3},
\begin{equation}
F ( \phi)= \int_{\Omega} \left[ \psi \left( \phi \right) +\frac{
\kappa
 }{2}\left | \nabla \phi\right |^2 \right] \,d\Omega,
 \label{eq:4}
\end{equation}
where $\Omega$ is the physical domain of the system, $\psi \left(
\phi \right)$ denotes the bulk free-energy density, and $k \left |
\nabla \phi\right |^2 /2$ accounts for the surface energy with a
positive coefficient $k$. If the system considered is a van der
Waals fluid, the bulk free energy has a double-well form
\cite{Jacqmin},
\begin{equation}
\psi(\phi)=\beta \phi^2(\phi-1)^2, \label{eq:5}
\end{equation}
where $\beta$ is a constant dependent on the interfacial thickness
$W$ and the surface tension $\sigma$  \cite{Jacqmin},
\begin {equation}
W= \sqrt{\frac{8\kappa}{\beta}} \label{eq:6}
\end{equation}
and
\begin{equation}
\sigma=\frac{\sqrt{2\kappa \beta}}{6} . \label{eq:7}
\end{equation}
The equilibrium interface profile can be obtained by minimizing
$F(\phi)$ with respect to the function $\phi$ , i.e., $\mu=0$ . In a
plane interface at the equilibrium condition, the order parameter
profile across the interface (along the $z$ direction) is
represented by \cite{Yan}
\begin{equation}
\phi(z)=\frac{1}{2}+\frac{1}{2}\tanh \left( \frac{2z}{W} \right).
\label{eq:9}
\end{equation}

In the present work, we will focus on the following
quasi-incompressible phase-field system, and the governing equations
are described as
\begin{subequations}
\begin{equation}
\frac{\partial \phi}{\partial t}+\nabla \cdot \phi \textbf{u}=\nabla
\cdot M_{\phi} \left( \nabla \mu \right),
\label{eq:30}
\end{equation}
\begin{equation}
\nabla \cdot \textbf{u} = S_1,
 \label{eq:31}
\end{equation}
\begin{equation}
\begin{split}
\frac{\partial (\rho \textbf{u})}{\partial t} + \nabla \cdot
\left(\rho \textbf{u} \textbf{u}\right) = & - \nabla \emph{p}+\nabla
\cdot \tau+\textbf{F},
\end{split}
\label{eq:32}
\end{equation}
\label{eq:32a}
\end{subequations}
where $\mathbf{F}$ represents the total force which is defined as
$\mathbf{F}_s+\mathbf{G}$, and $\mathbf{G}$ is the body force,
$\mathbf{F}_s$ is the surface tension with the potential form
$\mathbf{F}_s=\mu \nabla \phi$ \cite {Jacqmin,Zheng} if not
specified. Eq. (\ref{eq:31}) can be derived from Eq. (\ref{eq:1}),
if $S=\mathbf{u} \cdot \nabla \rho+\rho S_1$, and
$\tilde{\rho}=\rho_0$.

\subsection{The LB model for quasi-incompressible phase-field system}
\subsubsection{LB model for the N-S equations}
Based on the generalized LB model for fluid flow system, one can get
the equilibrium and force distribution function for the N-S
equations [Eq. ({\ref{eq:31}}) and Eq. ({\ref{eq:32}})] when
substituting $\tilde{\rho}=\rho_0$,  and $S=\mathbf{u} \cdot \nabla
\rho+\rho S_1$ into Eq. (\ref{eq:25}) and Eq. (\ref{eq:27}),
\begin{equation}
g_i^{eq}=\left\{
\begin{array}{l}
\rho_0+\frac{p}{c_s^2}(\omega_i-1)+\rho s_i(\mathbf{u}), \;\;\;i=0,\\
\frac{p}{c_s^2}\omega_i+\rho s_i(\mathbf{u}),
\;\;\;\;\;\;\;\;\;\;\;\;\;\;\;\;\;\;   i \ne 0,
\end{array}
\right. \label{eq:33}
\end{equation}
\begin{equation}
G_i=\omega_i\left[ \mathbf{u} \cdot \nabla \rho+\rho
S_1+\frac{\mathbf{c}_i \cdot \mathbf{F}}{c_s^2}+\frac{(\mathbf{c}_i
\mathbf{c}_i-c_s^2\mathbf{I}):\left(\mathbf{u \tilde{F}}+\mathbf{
\tilde{F}u}+\frac{2}{d}\rho c_s^2 S_1
\mathbf{I}\right)}{2c_s^4}\right ], \label{eq:34}
\end{equation}
where we take the bulk viscosity $\xi$ equal to $0$, and the dynamic
pressure satisfies $\delta p=O(Ma^2)$ so that the term of
$\partial_t p$ is neglected in the limit of a low Mach number.
Actually, the relationships of $\mathbf{u}=O(Ma)$ and
$\mathbf{F}=O(Ma)$ are also true. Specially, the force distribution
function $G_0$ can be simplified as
\begin{equation}
\begin{split}
G_0 & =\omega_0 \left(\mathbf{u}\cdot \nabla \rho+\rho S_1-\frac{2\mathbf{u}\cdot \tilde{\mathbf{F}}+\frac{2}{d} \rho c_s^2 S_1}{2c_s^2} \right) \\
&=\omega_0\left[ \rho S_1-\frac{1}{c_s^2}\left(\mathbf{u}\cdot
\mathbf{F}-\mathbf{u}\cdot \nabla p+\frac{1}{d}\rho c_s^2 S_1
\right) \right] \\
&=\omega_0 \left\{(1-\frac{1}{d})\rho
S_1-\frac{1}{c_s^2}[\mathbf{u}\cdot (\mathbf{F}- \nabla p)]\right
\}.
\end{split}
\end{equation}
From the above equation, one can find that $G_0$ is related to $S_1$
since the term $\mathbf{u}\cdot (\mathbf{F}- \nabla p)$ is the order
of $O(Ma^2)$, and is nonzero once $S_1 \ne 0$, thus the term $\Delta
t (\tau_g-\frac{1}{2}) G_0$ in Eq. (\ref{eq:28}) cannot be ignored
in the pressure expression.

We would like to point out that Eq. (\ref{eq:32a}) can reduce to the
 incompressible phase-field model if $S_1=0$.
However, based on Eqs. (\ref{eq:30}), (\ref{eq:31}), (\ref{eq:3aa})
and $S_1=0$, one can obtain \cite{Yang}
\begin{equation}
\frac{\partial \rho}{\partial t}+\nabla \cdot (\rho
\mathbf{u})=\frac{\rho_A-\rho_B}{\phi_A-\phi_B}  \nabla \cdot
[M_{\phi} \nabla \mu]. \label{eq:49aaa}
\end{equation}
It is clear that the mass conservation is constrained by the term on
the right hand of Eq. (\ref{eq:49aaa}), which is nonzero in the
interfacial region as long as $\rho_A \ne \rho_B$. Therefore, in the
incompressible phase-field model, the mass is not locally conserved.

To conserve mass locally,  Shen \emph{et al.} \cite{Shen} proposed a
quasi-incompressible  phase-field model. Subsequently, based on the
quasi-incompressible phase-field model, Yang \emph{et al.}
\cite{Yang} designed a lattice Boltzmann model for binary fluids.
Actually, Eq. (\ref{eq:32a}) can also reduce to quasi-incompressible
phase-field model in Ref. \cite{Shen} when $S_1=-\gamma \nabla \cdot
(M_{\phi} \nabla \mu)$ with $\gamma=\frac{\rho_A-\rho_B}{\phi_A
\rho_B-\phi_B \rho_A}$. It should also be noted that the momentum
equation (\ref{eq:32}) is different from those used in Refs.
\cite{Liang,Yang}, where the terms of $\mathbf{u}S$ and
$\left[\left(\xi-\frac{2}{d}\right) \mu \nabla \cdot \mathbf{u}
\right] \mathbf{I}$ are not included.

\emph{Remark 1.} Here we also give a compasion between the
generalized LB model and the one in Ref. \cite{Yang}. If we take
$S=c_s^2\left[\mathbf{u}\cdot \nabla \rho-\rho \gamma
\nabla\cdot(M_{\phi}\nabla \mu) \right]$,  and
$\tilde{\rho}=\frac{p}{c_s^2}$, the equilibrium distribution
function for the N-S equations can be written as
\begin{equation}
g_i^{eq}=\omega_i \left[ \frac{p}{c_s^2}+\rho s_i
(\mathbf{u})\right]
 \label{eq:35}
\end{equation}

Based on the following fact,
\begin{equation}
\frac{p}{c_s^2}=\sum_i g_i^{eq}=\sum_i g_i+\frac{\Delta t}{2}\sum_i
G_i, \sum_i G_i=S,
\end{equation}
the pressure can be calculated as
\begin{equation}
p=\sum_i c_s^2 g_i+\frac{\Delta t}{2} c_s^2\left[ \mathbf{u}\cdot
\nabla \rho -\rho \gamma\nabla \cdot (M_{\phi}\nabla \mu) \right],
\end{equation}
and the fluid velocity can be obtained from Eq. (\ref{eq:27a}). The
above derivation shows that our generalized flow field LB model can
reduce to Yang's model once the force distribution function $G_i$
takes the same form.
\subsubsection{LB model for the C-H equation}
The evolution
equation for the C-H equation can be given as
\begin{equation}
\begin{split}
h_i(\textbf{x}+\textbf{c}_i \Delta t,t+\Delta t)-h_i(\textbf{x},t)=&
-\frac{1}{\tau_{h}} \left[ h_i(\textbf{x},t)-h_i^{eq}(\textbf{x},t)
\right ]+\Delta t R_i(\textbf{x},t),
\end{split}
\label{eq:10}
\end{equation}
where $h_i(\textbf{x},t)$ is the distribution function of order
parameter $\phi$, $\tau_{h}$ is the non-dimensional relaxation time
related to the mobility, $h_i^{eq}(\textbf{x},t)$ is the local
equilibrium distribution function, which is defined as
\cite{Fakhari2010,Huang,Chai}
\begin{equation}
h_i^{eq}(\textbf{x},t)=\left \{
\begin{array}{l}
\phi+(\omega_i-1)\eta \mu, \;\;\;i=0\\
\omega_i \eta \mu+\omega_i \frac{\textbf{c}_i \cdot \phi
\textbf{u}}{c_s^2}, \;\;\;\;   i \ne 0,
\end{array}
\right. \label{eq:11}
\end{equation}
where $\eta$ is an adjustable parameter that controls the mobility.
$R_i(\textbf{x},t)$ is the source term and is given by \cite{Liang}
 \begin{equation}
 {R_i}=\left(1-\frac{1}{2\tau} \right) \frac{\omega_i \mathbf{c}_i \cdot \partial_t \phi
 \textbf{u}}{c_s^2}.
 \label{eq:14}
 \end{equation}
In our simulations, the first-order explicit difference scheme
 is used to compute the time derivative in Eq.
(\ref{eq:14}),
\begin{equation}
\partial_t \phi \textbf{u}|_{(\mathbf{x},t)}=[\phi
\textbf{u}|_{(\mathbf{x},t)}-\phi \textbf{u}|_{(\mathbf{x},t-\Delta
t)} ]/{\Delta t}.
\end{equation}

Through the Chapman-Enskog analysis, the order parameter $\phi$ is
calculated by
\begin{equation}
\phi=\sum_i h_i, \label{eq:18}
\end{equation}
the C-H equation can be recovered with second-order accuracy and the
mobility can be determined by
\begin{equation}
M_{\phi}=\eta c_s^2 (\tau_h-0.5)\Delta t. \label{eq:17}
\end{equation}

In a two-phase system, the viscosity is no longer a uniform value
due to its jump at the interface. In this work, the following
viscosity with a inverse linear form \cite{Lee2,Fakhari2017} is
adopted to ensure a smooth viscosity across the interface if not
specified,
\begin{equation}
\frac{1}{\nu}=\frac{\phi-\phi_B}{\phi_A-\phi_B}(\frac{1}{\nu_A}-\frac{1}{\nu_B})+\frac{1}{\nu_B}.
\end{equation}


In addition, to determine the gradient terms in the source term
$G_i$, surface tension $\mathbf{F}_s$ and chemical potential $\mu$,
the following isotropic schemes are adopted to discretize the
first-order spatial derivative and the Laplacian operator
\cite{Liang}:
\begin{subequations}
\begin{equation}
\nabla \zeta(\mathbf{x},t)=\sum_{i \ne 0}\frac{ \omega_{i}
\mathbf{c}_{i} \zeta(\mathbf{x}+\mathbf{c}_{i} \Delta t,t)}{c_s^2
\Delta t}, \label{eq:30a}
\end{equation}
\begin{equation}
\nabla^2 \zeta(\mathbf{x},t)=\sum_{i\ne 0} \frac{2\omega_i[\zeta
(\mathbf{x}+\mathbf{c}_i \Delta t,t)-\zeta(\mathbf{x},t)]}{c_s^2
\Delta t^2}, \label{eq:30b}
\end{equation}
\label{eq:30c}
\end{subequations}
where $\zeta$ is an arbitrary function. It should be noted that the
schemes (\ref{eq:30c}) not only have a secondary-order accuracy in
space, but also can ensure the global mass conservation of a
multiphase system \cite{Liang1}.

\section{Model validation}
In this section, several examples including a spinodal
decomposition, a static droplet, layered Poiseuille flows and a
bubble rising flow, are performed to test our LB model for
incompressible ($S_1=0$) and quasi-incompressible [$S_1=-\gamma
\nabla \cdot (M_{\phi} \nabla \mu)$] two-phase flows. In our
simulations, the D$2$Q$9$ lattice structure is adopted. We attempt
to conduct a detailed comparison between the present and the
analytical solutions or some available results in each test.
\subsection{Spinodal decomposition}
\begin{figure}[h]
\subfigure[]{ \label{fig:mini:subfig:a0}
\includegraphics[width=1.0\textwidth]{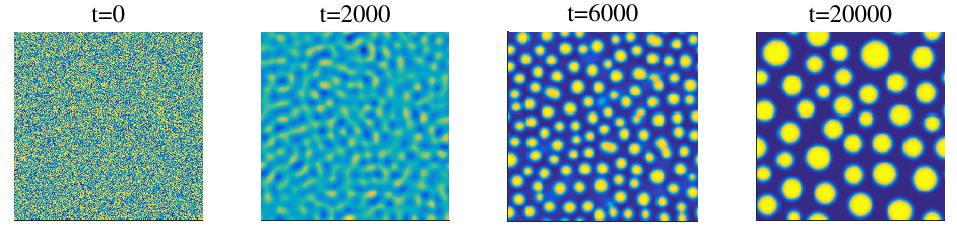}}
\subfigure[]{ \label{fig:mini:subfig:b0}
\includegraphics[width=1.0\textwidth]{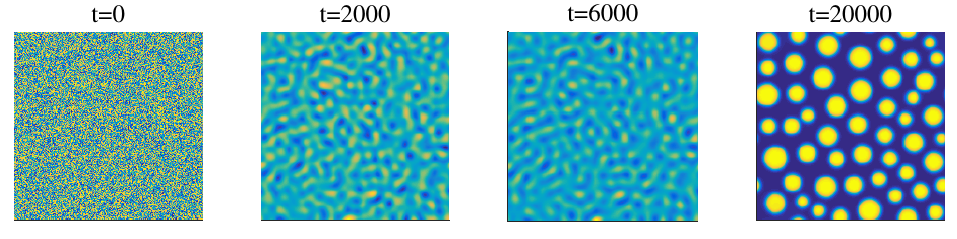}}
\caption{(Colour online) Separation of binary fluid: distribution of
order parameter with modified pressure expression Eq. (\ref{eq:28}),
(a) our quasi-incompressible model, (b) our incompressible model. }
\label{Spinodal_1}
\end{figure}

Spinodal decomposition \cite{Cahn1} is a mechanism for the rapid
unmixing of a fluid mixture with two different species. The spinodal
decomposition phenomenon will take place when the initial
homogeneous mixture is unstable in the presence of small
fluctuations. In this section, the separation of two emulsified
fluids is simulated with different pressure expressions, i.e., Eq.
(\ref{eq:28}) and Eq. (\ref{eq:29}), and this example is mainly used
to demonstrate the differences in calculation results obtained by
present incompressible and quasi-incompressible models. The
computational domain is $NY \times NX= 200\times 200$, the initial
distribution of order parameter with a small fluctuation is given by
\begin{equation}
\phi(x,y)=\frac{1}{3}+rand(x,y),
\end{equation}
where $rand(x,y)$ is a random function with the maximum amplitude of
$1\%$. The initial velocity is zero in the whole domain and the
periodic boundary conditions are applied at all boundaries. The
model parameters are fixed as
$\sigma=0.03,W=4,M_{\phi}=0.1,\phi_A=1,\phi_B=0,\rho_A
/\rho_B=5,\nu_A/\nu_B=0.1,\tau_h=1$. Fig. {\ref{Spinodal_1}} shows
the time evolution of the order-parameter distribution by our
incompressible model (IM) and quasi-incompressible model (QIM) with
 Eq. (\ref{eq:28}). It can be found from this figure
that small fluctuations of order-parameter gradually become larger,
then some small droplets are formed, and the diameters of the
droplets gradually become larger. Finally, the phenomenon of
fluid-fluid separation can be observed as expected. However, the
results of the incompressible and quasi-compressible models are
significantly different, which may be caused by the fact that the
term $\gamma \nabla \cdot (M_{\phi} \nabla \mu)$ has been neglected
in incompressible model, and thus mass conservation is not satisfied
locally.

\begin{figure}[h]
\subfigure[]{ \label{fig:mini:subfig:a1}
\includegraphics[width=0.8\textwidth]{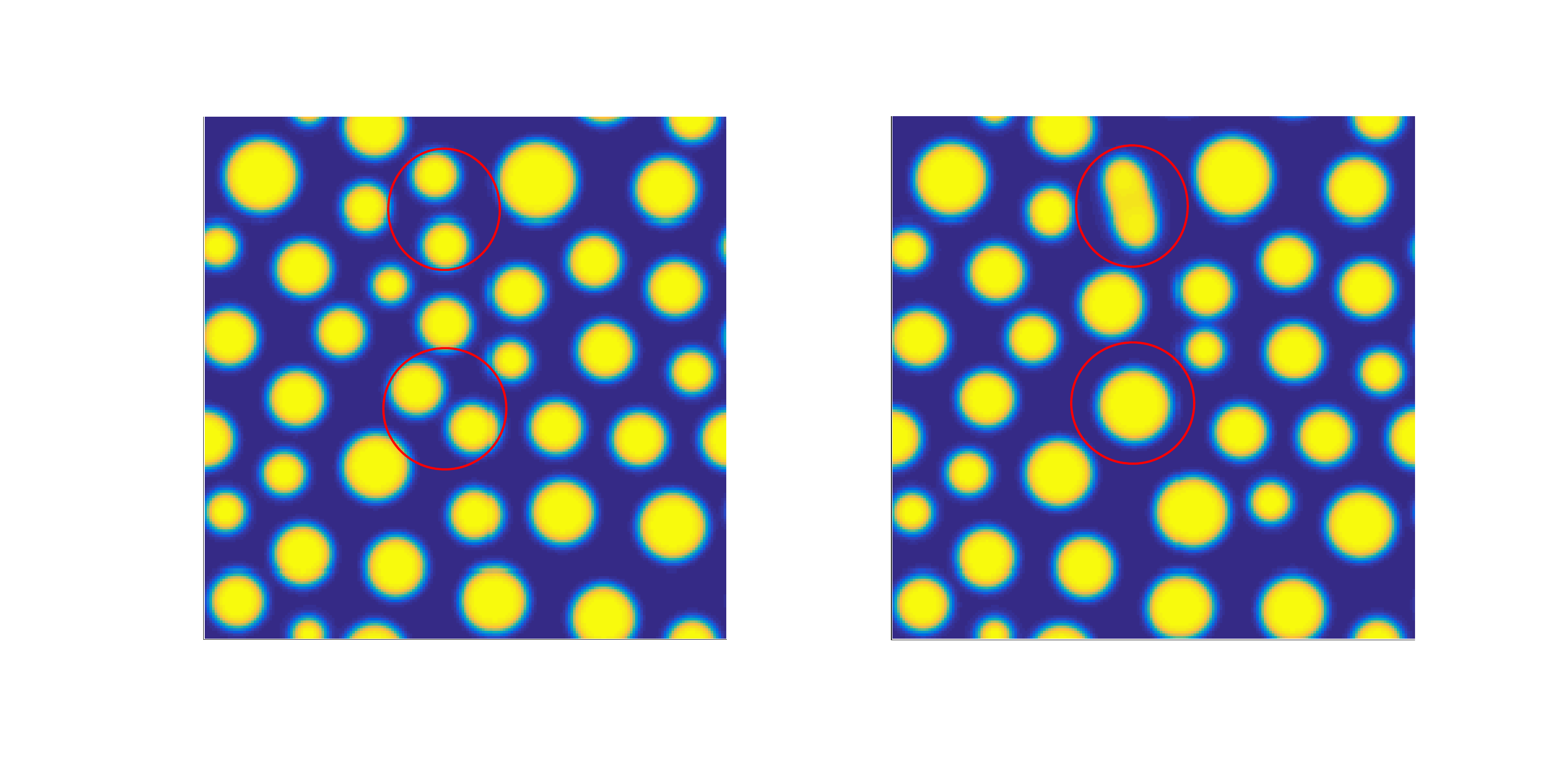}}
\subfigure[]{ \label{fig:mini:subfig:b1}
\includegraphics[width=0.8\textwidth]{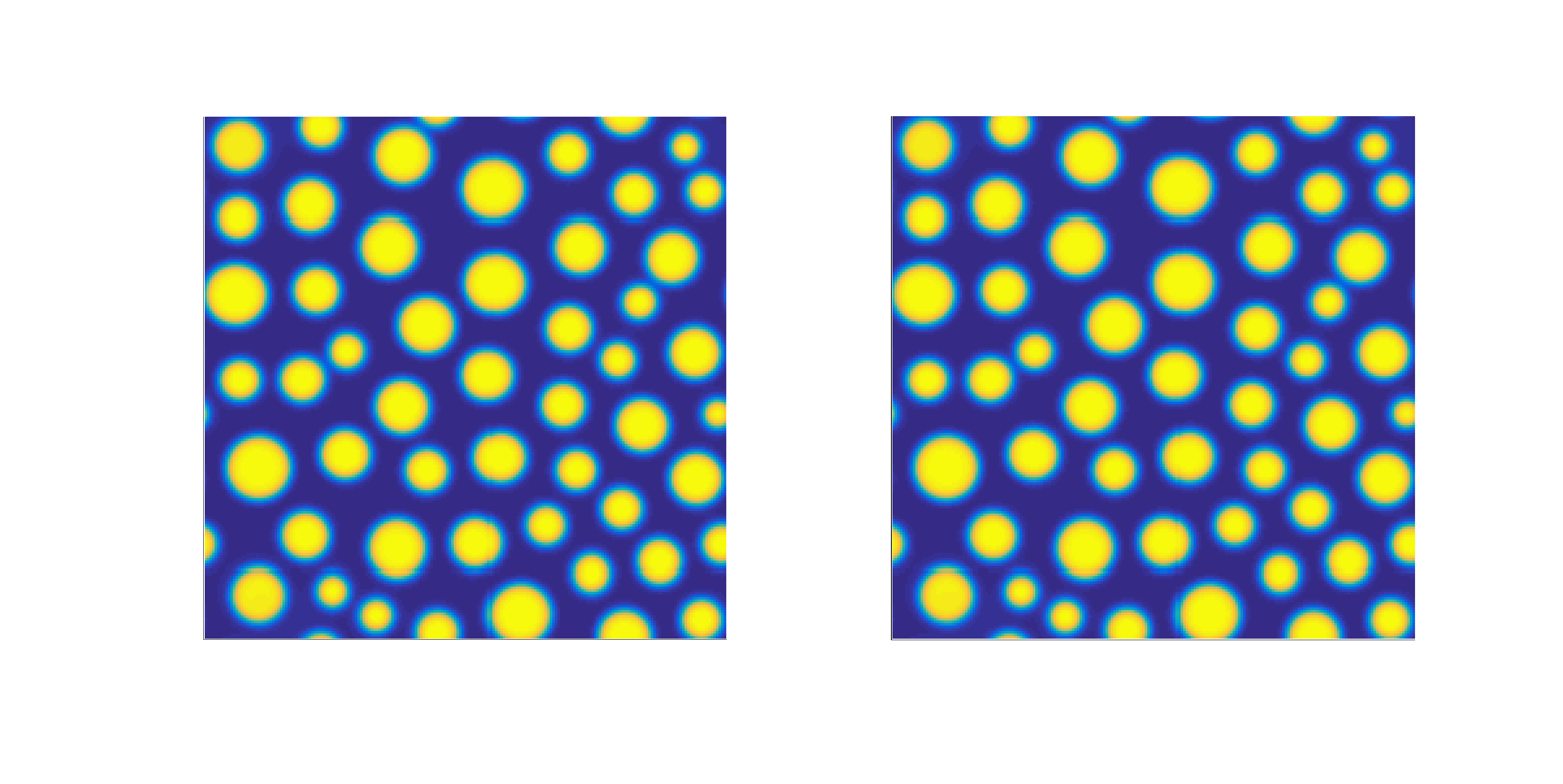}}
\caption{(Colour online) Distribution of order parameter with Eq.
(\ref{eq:28}) on the left, and Eq. (\ref{eq:29}) on the right, (a)
the quasi-incompressible model at t=20000, (b) the incompressible
model at t=20000. } \label{Spinodal_2}
\end{figure}

To illustrate the difference between two pressure expressions, we
presented some results in Fig. {\ref{Spinodal_2}}. As seen in Fig.
{\ref{Spinodal_2}}(a), the distributions of order parameter within
the red circles are distinctly different, which means that the term
$\Delta t (\tau_g-0.5)G_0$ in Eq. (\ref{eq:28}) has a significant
influence on the numerical results. However, for incompressible LB
model, there are no apparent differences, which is due to the fact
that the effect of $\Delta t (\tau_g-0.5)G_0$ can be ignored when
$S_1=0$, and $G_0$ is the order of $O(Ma^2)$. Based on the above
results, one can conclude that the pressure expression Eq.
(\ref{eq:28}) is more general, and would be adopted in the following
simulations.

\subsection{Static droplet}
\begin{figure}[h]
\subfigure[]{ \label{fig:mini:subfig:a3}
\includegraphics[width=2.5in]{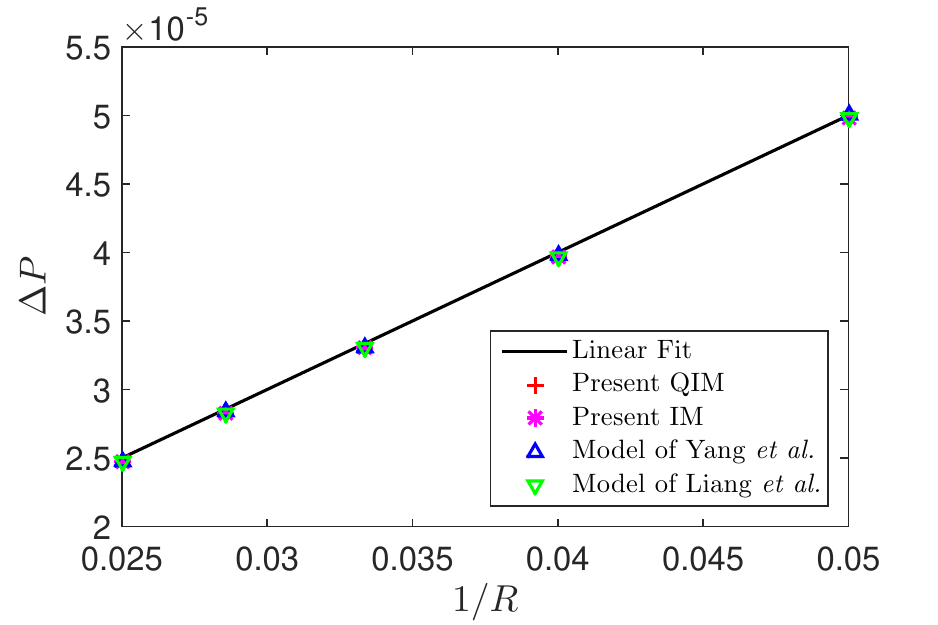}}
\subfigure[]{ \label{fig:mini:subfig:b3}
\includegraphics[width=2.5in]{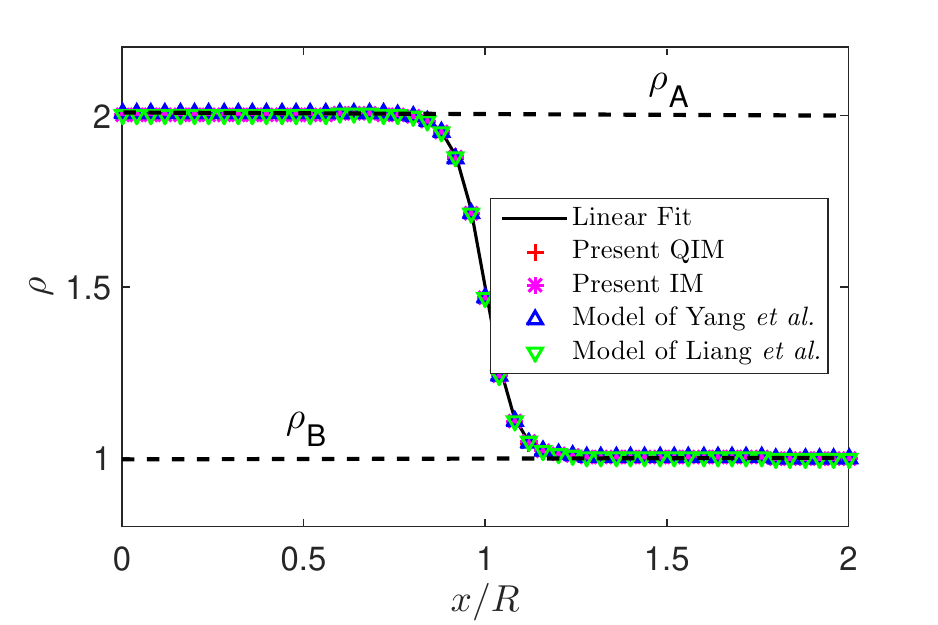}}
\subfigure[]{ \label{fig:mini:subfig:c3}
\includegraphics[width=2.5in]{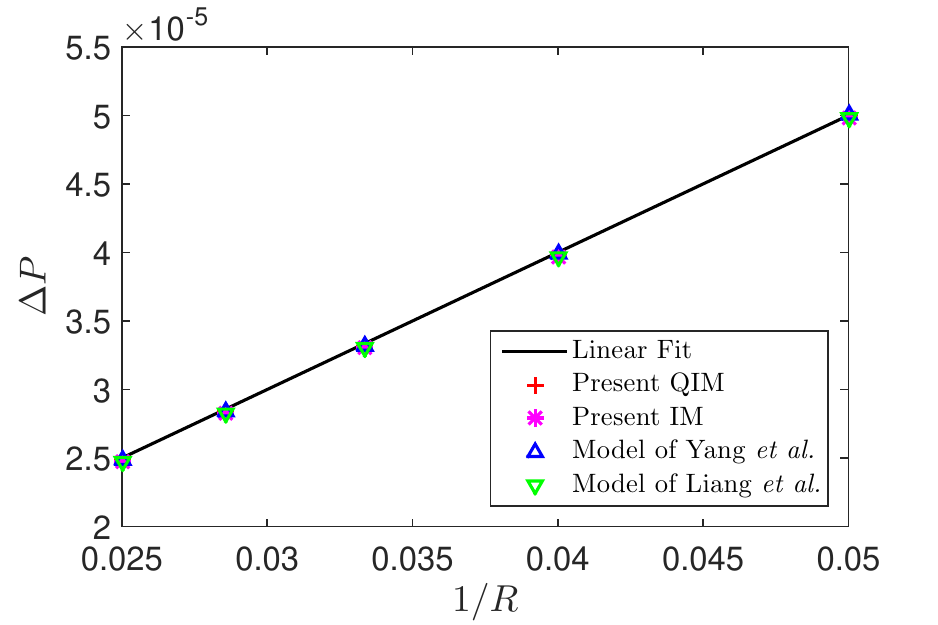}}
\subfigure[]{ \label{fig:mini:subfig:d3}
\includegraphics[width=2.5in]{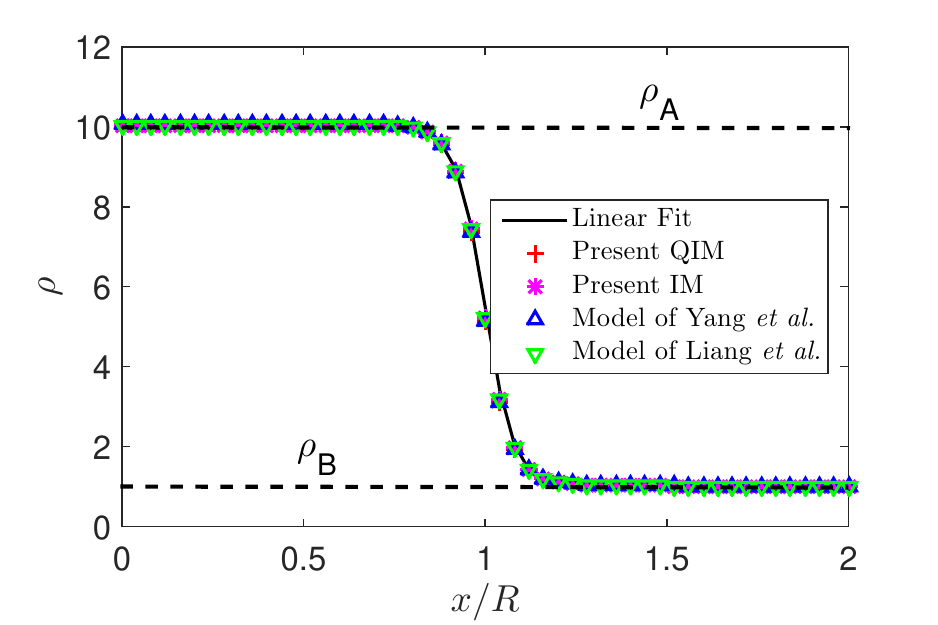}}
\subfigure[]{ \label{fig:mini:subfig:e3}
\includegraphics[width=2.5in]{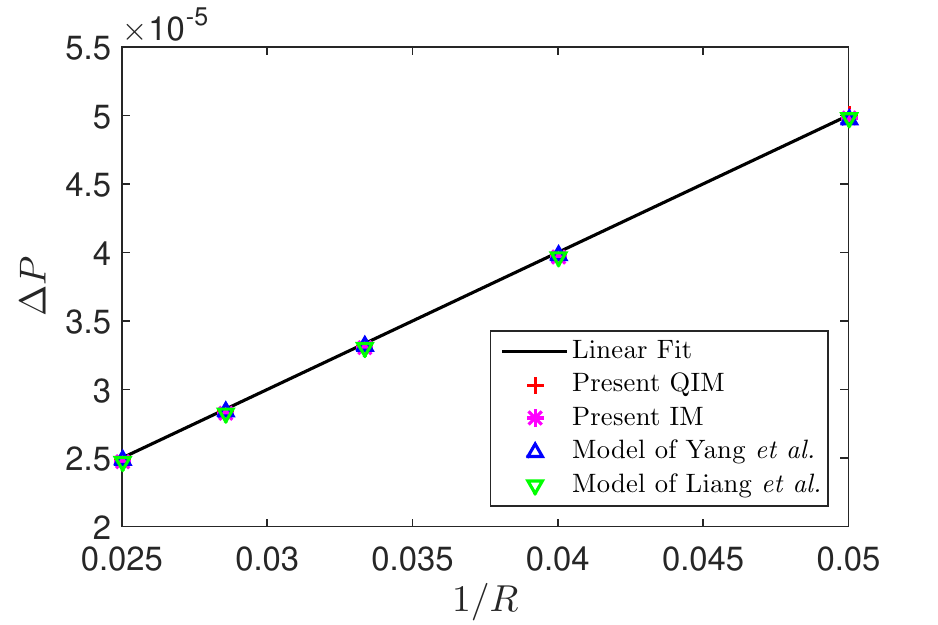}}
\subfigure[]{ \label{fig:mini:subfig:f3}
\includegraphics[width=2.5in]{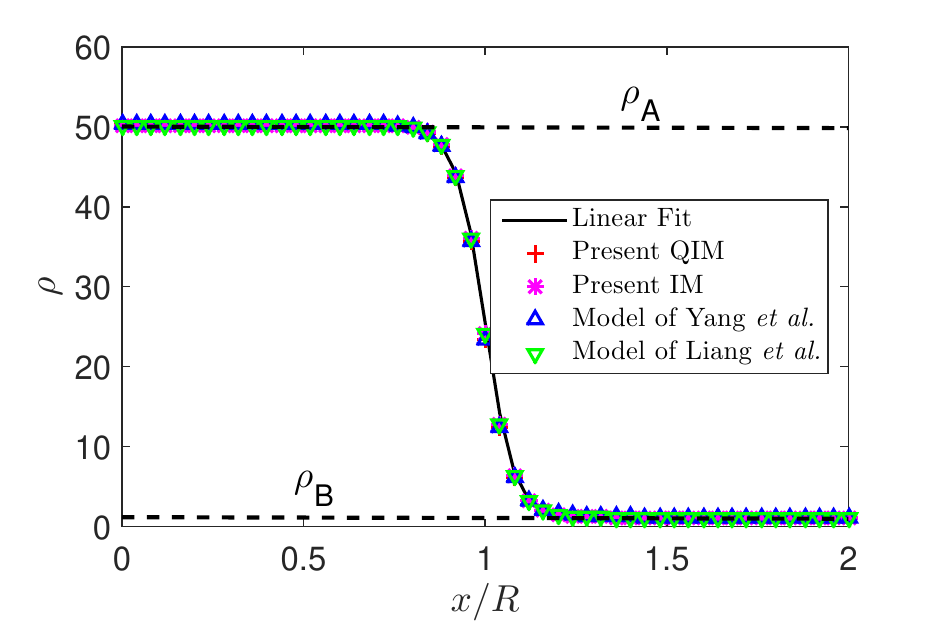}}
\caption{(Color online) Tests of Laplace' law [(a), (c), (e)] and
density profiles [(b), (d), (f)] at different density ratios
($\rho_A/\rho_B=2, 10, 50$), where the droplet radius is 25.}
 \label{Static1}
\end{figure}

A $2D$ static droplet is popular problem to verify LB models for
two-phase flow \cite{Zu,Ba,Liang,Lee1,Fakhari2010}. In this
subsection, we will consider this problem with different density
ratios. Initially, a circular droplet with the radius ranging from
$20$ to $40$ is placed in the middle of the computational domain
with $NX \times NY=100 \times 100$. The initial order parameter is
given by
\begin{equation}
\phi(x,y)=0.5+0.5\tanh
\frac{2\left[R-\sqrt{(x-50)^2+(y-50)^2}\right]}{W}, \label{eq:36}
\end{equation}
where $R$ is the droplet radius, and surface tension is expressed as
$\mathbf{F_s}=-\phi \nabla \mu$. In numerical simulations, we set
the density ratio to be $\rho_A/\rho_B=2,10$, and $50$.  The other
physical parameters are fixed as $\rho_B=1, \tau_h=1, \tau_g=0.8,
\sigma=0.001,M_{\phi}=0.02$, and the periodic boundary conditions
are applied at all boundaries. We first verify the LB model with the
well-known Laplace's law,
\begin{equation}
\Delta P=\frac{\sigma}{R},
\end{equation}
where $\Delta P$ is the pressure jump across the interface, $R$ is
the radius of the droplet. $P$ is calculated by $P=p_0-\kappa \phi
\nabla^2\phi+\kappa|\nabla \phi|^2/2+p$ with the equation of state
$p_0=\phi\partial_{\phi}\psi-\psi$ \cite{Zu,ZhengL}. We performed
some simulations and presented the results in Fig. {\ref{Static1}}.
From Fig. {\ref{Static1}}(a), {\ref{Static1}}(c), {\ref{Static1}}(e)
one can find that the results of present models and those in Refs.
\cite{Liang,Yang} agree well with the Laplace law. In order to show
the difference between these models more clearly, we also give the
relative errors of pressure jump with the density ratio
$\rho_A/\rho_B=50$ in Table {\ref{Tab_Exp2_1}}. In general, the
present quasi-incompressible LB model and quasi-incompressible model
in Ref. \cite{Yang} are more accurate than incompressible LB models.
This is because that quasi-incompressible model is physically more
reasonable than incompressible model. Besides, it is also found that
usually present QIM is more accurate than the model in Ref.
\cite{Yang}.

\begin{figure}[h]
\centering
\includegraphics[width=0.6\textwidth]{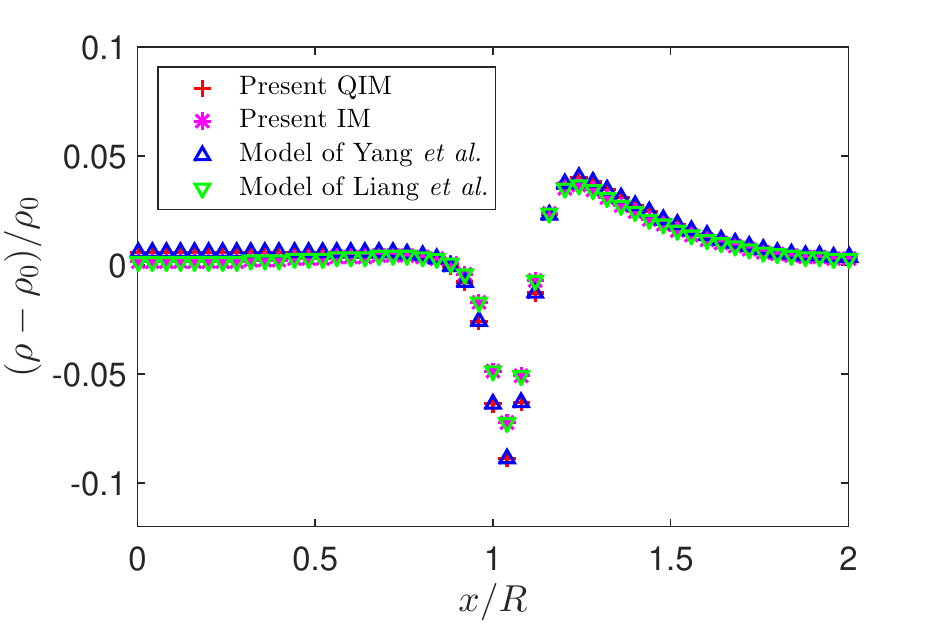}
\caption{(Color online) Tests of diffusion errors of density
profiles with $\rho_A/\rho_B=10$ and $R=25$.}
 \label{Static2}
\end{figure}

In addition, we also presented the density profiles along  the
horizontal center line at $t=5\times 10^{5}\Delta t$  in
 Figs. {\ref{Static1}}(b), {\ref{Static1}}(d), and
 {\ref{Static1}}(f). From these figures, one can observe that all the numerical results are very close to
  the analytical solutions given by Eq. (\ref{eq:36}). To
  give a quantitatively estimation on the accuracy of numerical results, we also
 measured the relative errors of density,
  i.e., [$(\rho-\rho_0)/\rho_0$] with $\rho_A/\rho_B=10$ in Fig.
  {\ref{Static2}}(a), where $\rho_0$ is the analytical solution.
  Different from the results of Laplace's law, we can find that the
 present incompressible LB models and the one in Ref. \cite{Liang} produce
  smaller errors than quasi-incompressible models, which means that the
quasi-incompressible and incompressible models each have their own
advantages.

\begin{table}[tbp]
\caption{Relative error of pressure jump with the density ratio
$\rho_A/\rho_B=50$.} \label{Tab_Exp2_1} \centering
\begin{tabular}{lcccccc}
\hline\hline
$R$  &&  & Present QIM & Present IM&Yang \emph{et al.} &Liang \emph{et al.} \\
\midrule[1pt]
20 && & $0.02\%$ & $0.36\%$ & $0.50\%$ & $0.36\%$ \\
25 && & $0.45\%$ & $0.73\%$ & $0.63\%$ & $0.73\%$  \\
30 && & $0.42\%$ & $0.84\%$ & $0.63\%$ & $0.84\%$ \\
35 && & $0.53\%$ & $0.95\%$ & $0.60\%$ & $0.95\%$ \\
40 && & $0.68\%$ & $1.00\%$ & $0.48\%$ & $1.00\%$ \\
\hline\hline
\end{tabular}
\end{table}

\subsection{Layered Poiseuille flow}
The layered Poiseuille flow between two parallel plates is also a
classical two-phase problem which provides a good benchmark for
validating the LB models
\cite{Ba,Zu,Wang,Fakhari2017,Gan,Liang2018}. Considering a channel
flow of two immiscible fluids driven by a const pressure gradient
$G$ in the flowing direction ($x$-direction). Initially, fluid $A$
is located in the upper region of the channel ($0<y \leq h$), while
fluid $B$ is at the bottom region ($-h \leq y \leq 0$). When the
flow is sufficiently slow and no instabilities occur at the
interface, an steady analytical solution of velocity can be
obtained,
\begin{equation}
\mathbf{u}_{x,a}(y)=\left \{
\begin{array}{l}
\frac{Gh^2}{2\mu_A}\left[-\left(\frac{y}{h}\right)^2-\frac{y}{h}\left(\frac{\mu_A-\mu_B}{\mu_A+\mu_B} \right)+\frac{2\mu_A}{\mu_A+\mu_B} \right], \;\;\;\;\;0<y \leq h\\
\frac{Gh^2}{2\mu_B}\left[-\left(\frac{y}{h}\right)^2-\frac{y}{h}\left(\frac{\mu_A-\mu_B}{\mu_A+\mu_B} \right)+\frac{2\mu_B}{\mu_A+\mu_B} \right]. \;\;\;-h \leq y \leq 0\\
\end{array}
\right. \label{eq:37}
\end{equation}
In this work, $G$ is given as $G=u_c(\mu_A+\mu_B)/h^2$, and to
ensure the stability of the interface, $u_c$ is fixed as
$u_c=5\times 10^{-5}$. To quantitatively describe the accuracy of
the present models and also convenient compare with the existing LB
models, the following relative error is adopted \cite{Liang2018},
\begin{equation}
Error=\frac{\sum\limits_y |u_x^n(y,t)-u_x^a(y)|}{\sum \limits_y
|u_x^a(y)|},
\end{equation}
where the subscripts $a$ and $n$ denote the analytical and numerical
solutions.

\begin{figure}[h]
\subfigure[]{ \label{fig:mini:subfig:a4}
\includegraphics[width=2.5in]{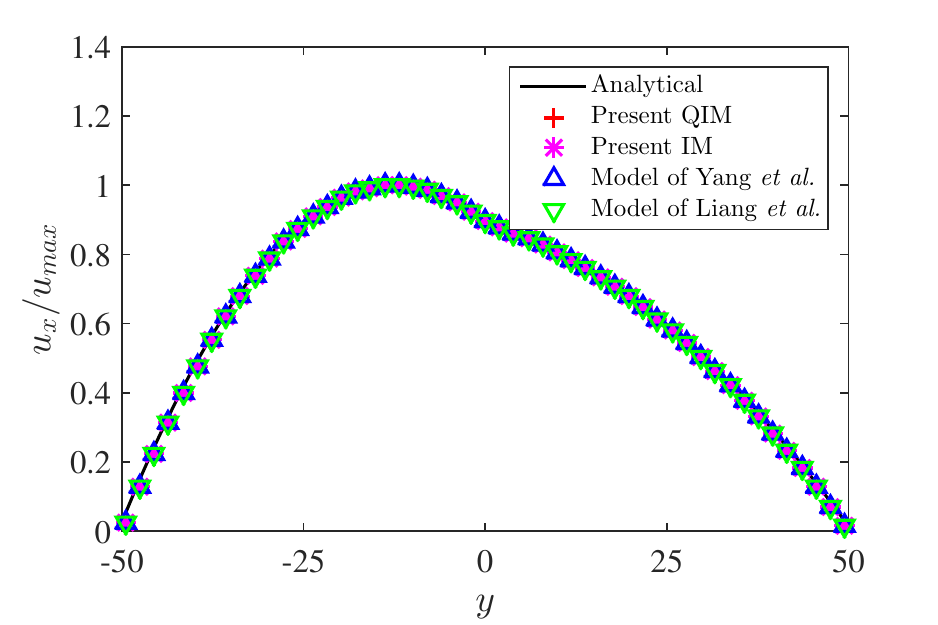}}
\subfigure[]{ \label{fig:mini:subfig:b4}
\includegraphics[width=2.5in]{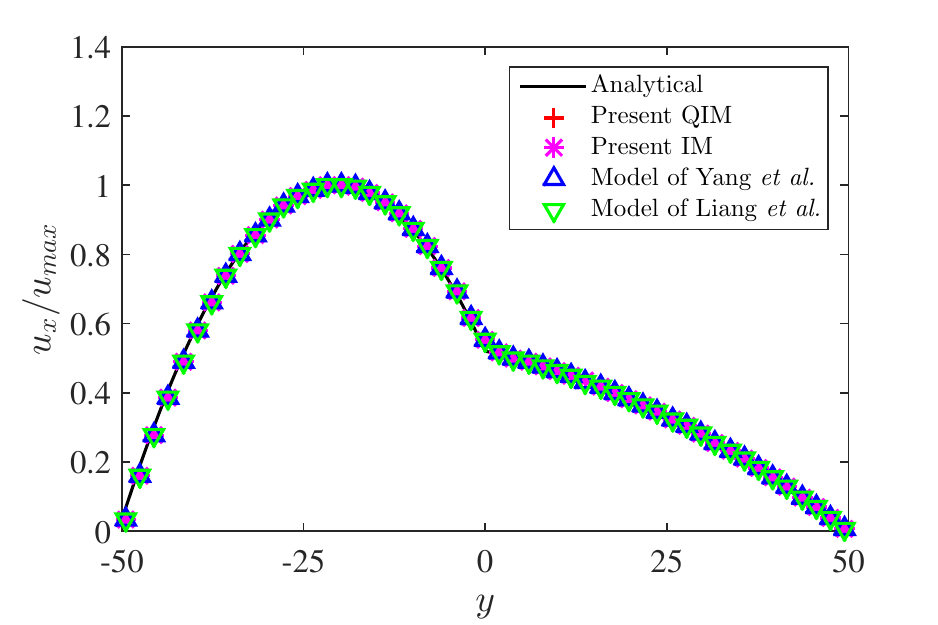}}
\subfigure[]{ \label{fig:mini:subfig:c4}
\includegraphics[width=2.5in]{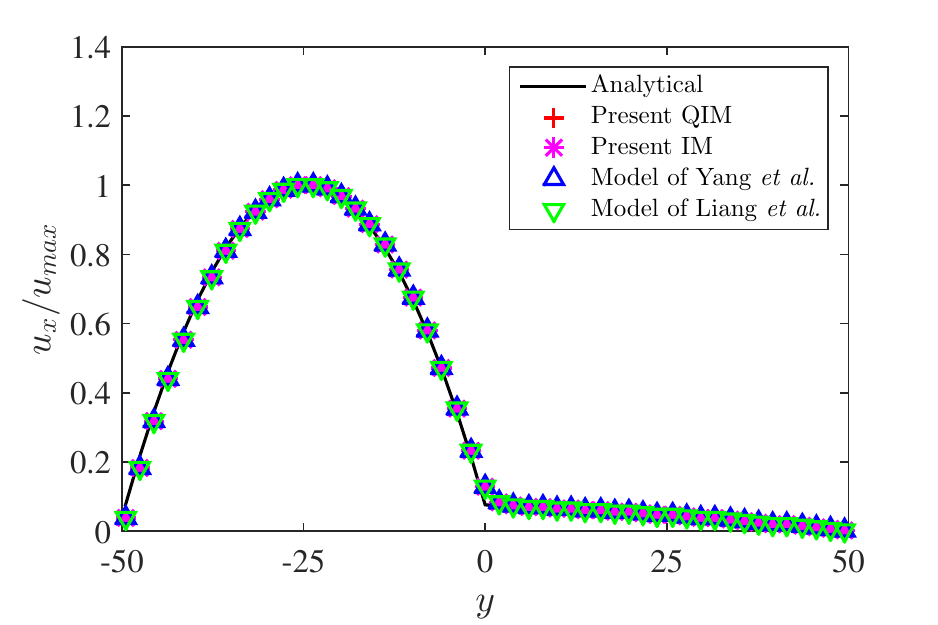}}
\subfigure[]{ \label{fig:mini:subfig:d4}
\includegraphics[width=2.5in]{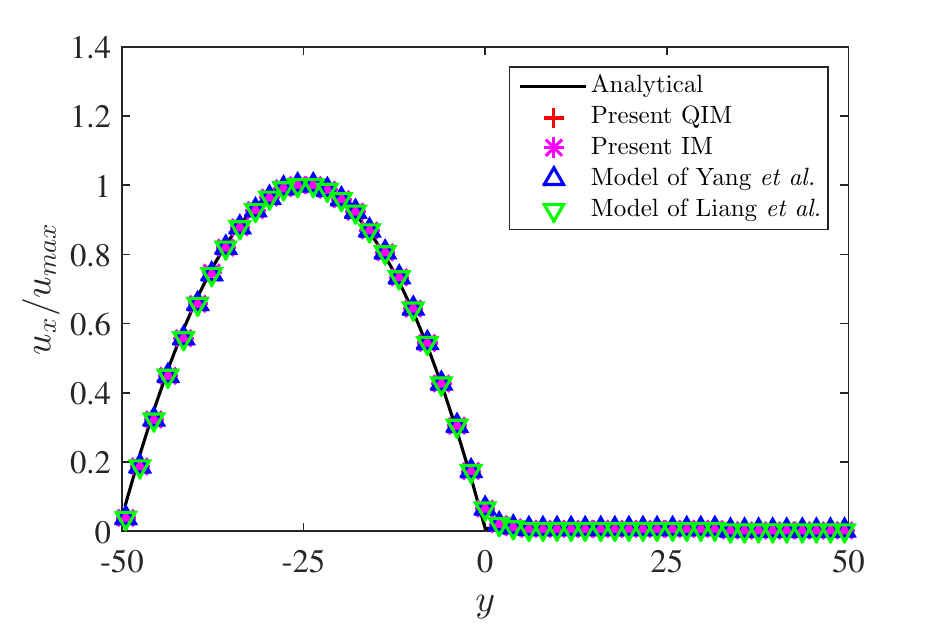}}
\caption{(Color online) Comparison of the velocity distributions
obtained by present QIM, IM, model of Yang \emph{et al.}
\cite{Yang}, and model of Liang \emph{et al.} \cite{Liang} with the
corresponding analytical solutions (solid line): (a)
$\mu_A/\mu_B=3$, (b) $\mu_A/\mu_B=10$, (c) $\mu_A/\mu_B=100$, (d)
$\mu_A/\mu_B=1000$. $u_x$ is normalized by he maximum speed of
analytical solution $u_{max}$. }
 \label{Layered1}
\end{figure}

In our simulations, the computational domain is chosen as $NY\times
NX=100\times10$. Periodic boundary conditions are applied in the
$x$-direction, and the halfway bounce-back boundary conditions are
enforced on the top and bottom walls. We first investigated the
effects of viscosity ratio. To this end, four different cases with
viscosity ratios of $\mu_A/\mu_B=3,10,100, 1000$ are considered. The
other parameters are given as $W=4$, $\sigma=0.001$,
$M_{\phi}=0.1$, $\rho_l/\rho_g=1$. 
Based on the results in Fig. {\ref{Layered1}}, one can find that the
numerical results of present LB models and some others
\cite{Yang,Liang} agree well with the analytical solutions for
different viscosity ratios. We also calculated the relative errors
of velocity under different viscosity ratios, and the results are
given in Table {\ref{Tab_Exp3_1}}. From this table, one can observe
that the relative error increases as the viscosity ratio becomes
larger, and for a fixed viscosity ratio, the relative errors of all
these models are almost the same. 
This is because if the density ratios are equal to $1$, which leads
to $\gamma=0$, the quasi-incompressible and incompressible models
are equivalent except for some terms with the order of $O(Ma^2)$.

\begin{figure}[h]
\subfigure[]{ \label{fig:mini:subfig:a5}
\includegraphics[width=2.5in]{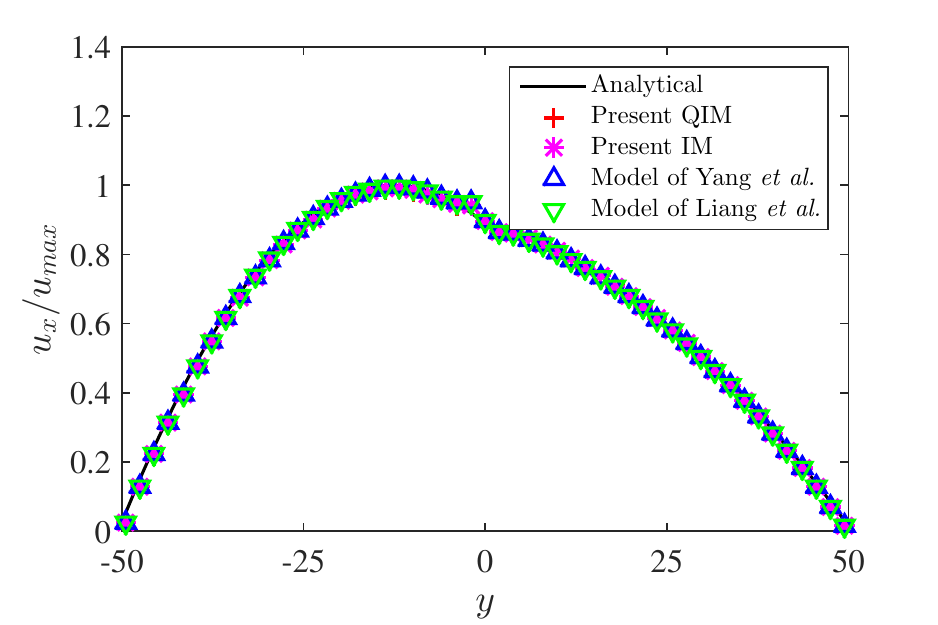}}
\subfigure[]{ \label{fig:mini:subfig:b5}
\includegraphics[width=2.5in]{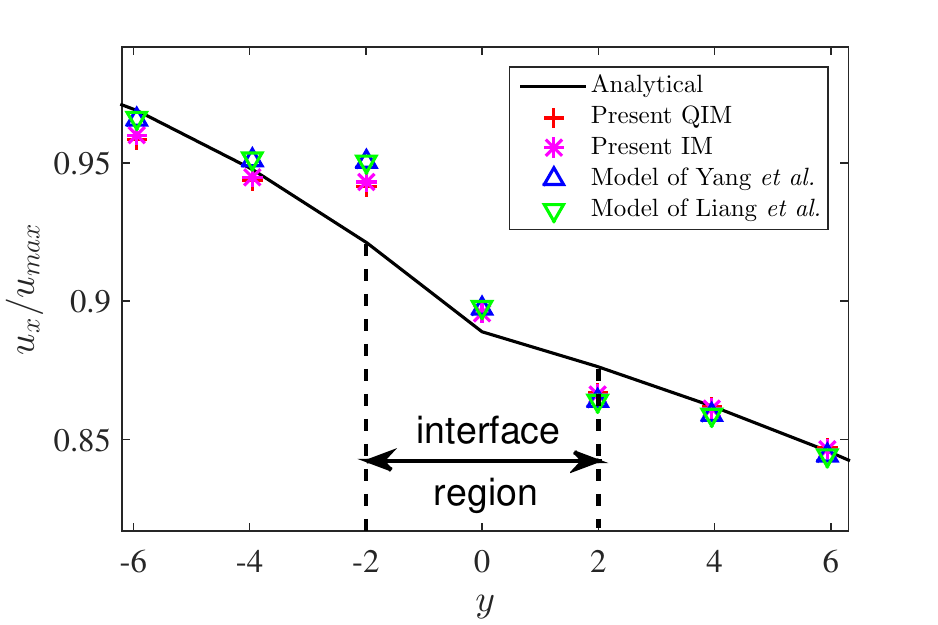}}
\caption{(Color online) Comparison of the velocity distributions
obtained by present QIM, IM, model of Yang \emph{et al.}
\cite{Yang}, and model of Liang \emph{et al.} \cite{Liang} with the
analytical solution (solid line), where $\rho_A/\rho_B=3$, and
$\mu_A/\mu_B=3$: (a) velocity distribution in the whole region, (b)
velocity distribution in the interface region. }
 \label{Layered2}
\end{figure}

We then simulated the layered Poiseuille flow with another density
ratio $\rho_A/\rho_B =3$, and viscosity ratio is equal to $3$.
 Here the dynamic viscosity is given by \cite{Ren,Liang2018}
\begin{equation}
\mu=\left \{
\begin{array}{l}
\mu_A,\;\;\;\;\;\phi \geq 0.5,\\
\mu_B,\;\;\;\;\;\phi < 0.5.\\
\end{array}
\right.
\end{equation}
The other parameters are the same as those stated above. From Fig.
{\ref{Layered2}}(a), one can find that there is  a good agreement
between the numerical solutions of the four LB models and the
analytical solutions except in the interface region. 
Fig. {\ref{Layered2}}(b) is an enlarged view of the interface region
in Fig. {\ref{Layered2}}(a). From this figure, it can be observed
that the present QIM and IM produce smaller errors than the models
of Yang \emph{et al.} \cite{Yang} and Liang \emph{et al.}
\cite{Liang} in the interface area, while their models are more
accurate in the bulk region. In other words, our models have a
better performance in the interface region, in contrast, the models
of Yang \emph{et al.} \cite{Yang} and Liang \emph{et al.}
\cite{Liang} are more accurate in the bulk region.

\begin{table}[tbp]
\caption{Relative errors of velocity for different viscosity
ratios.} \label{Tab_Exp3_1} \centering
\begin{tabular}{lcccccccccc}
\hline\hline
Models   &$\frac{\mu_A}{\mu_B}=3$ & $\frac{\mu_A}{\mu_B}=10$ &$\frac{\mu_A}{\mu_B}=100$ & $\frac{\mu_A}{\mu_B}=1000$\\
\midrule[1pt]
Present QIM                     & $1.04\%$ & $1.30\%$ & $1.90\%$ & $2.16\%$ \\
Present IM                      & $1.04\%$ & $1.30\%$ & $1.90\%$ & $2.16\%$ \\
Yang \emph{et al.} \cite{Yang}  & $1.04\%$ & $1.30\%$ & $1.90\%$ & $2.16\%$ \\
Liang \emph{et al.} \cite{Liang}& $1.04\%$ & $1.30\%$ & $1.90\%$ & $2.16\%$ \\
\hline\hline
\end{tabular}
\end{table}

\subsection{Bubble rising under buoyancy}
\begin{figure}[h]
\subfigure[]{ \label{fig:mini:subfig:a6}
\includegraphics[width=2.5in]{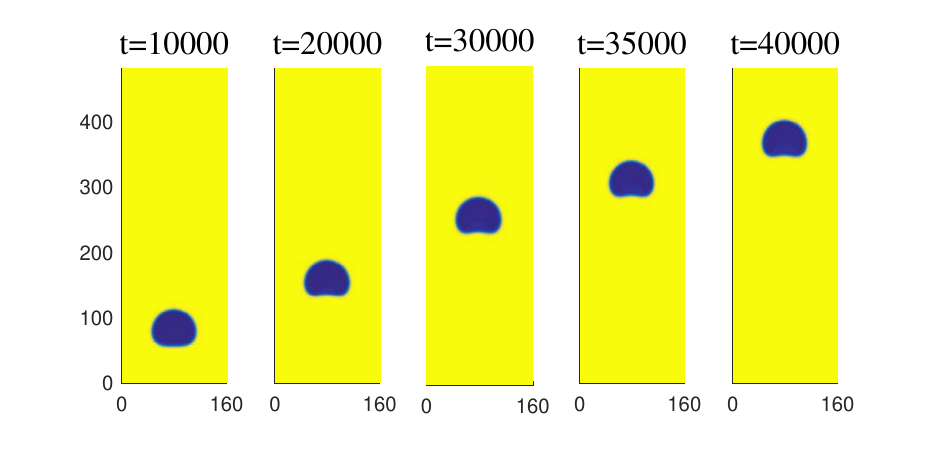}}
\subfigure[]{ \label{fig:mini:subfig:b6}
\includegraphics[width=2.5in]{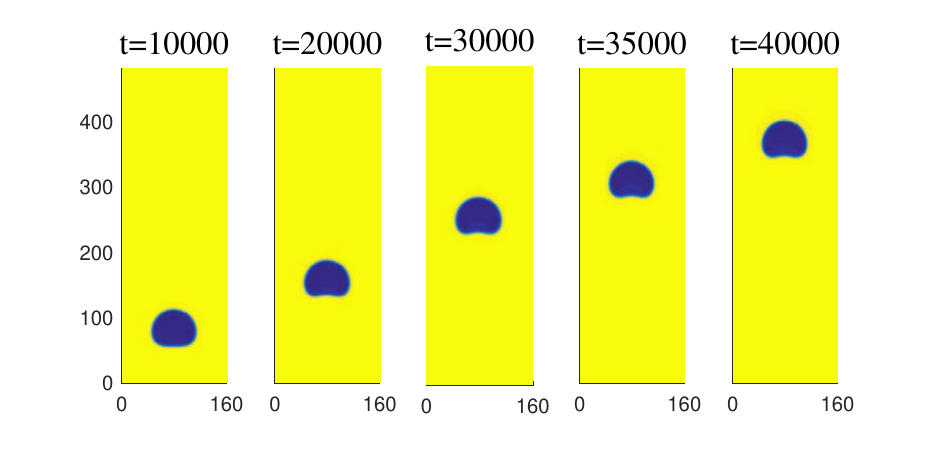}}
\subfigure[]{ \label{fig:mini:subfig:c6}
\includegraphics[width=2.5in]{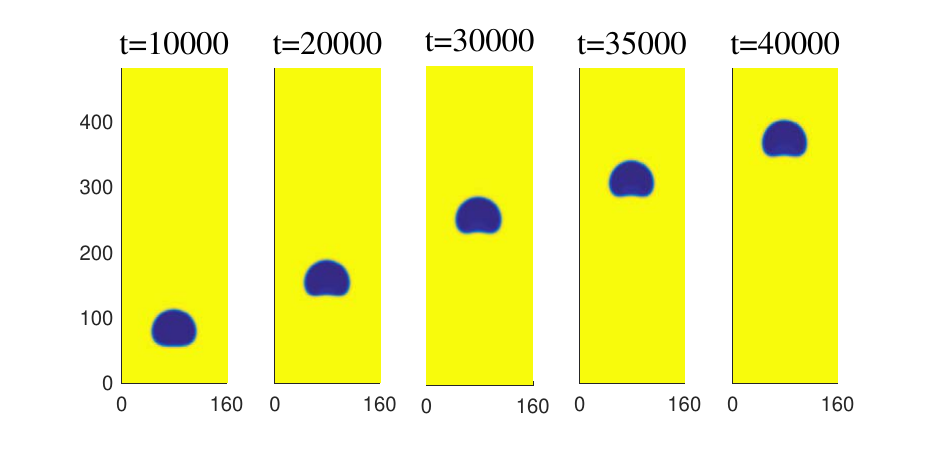}}
\subfigure[]{ \label{fig:mini:subfig:d6}
\includegraphics[width=2.5in]{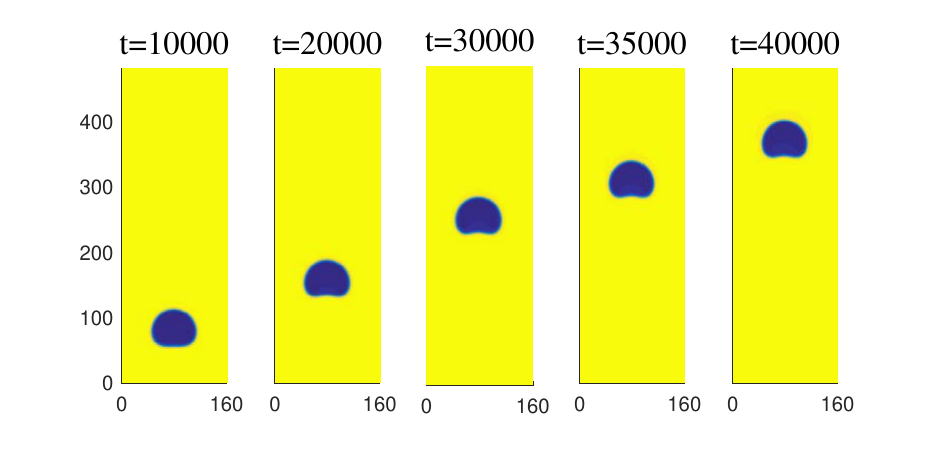}}
\caption{(Color online) Density distributions of the rising bubble
at $t=10000,20000,30000,35000,40000$, (a) present QIM, (b) present
IM, (c) model of Yang \emph{et al.} \cite{Yang}, and (d) model of
Liang \emph{et al.} \cite{Liang}.}
 \label{bubble1}
\end{figure}

\begin{figure}[h]
\centering
\includegraphics[width=1.0\textwidth]{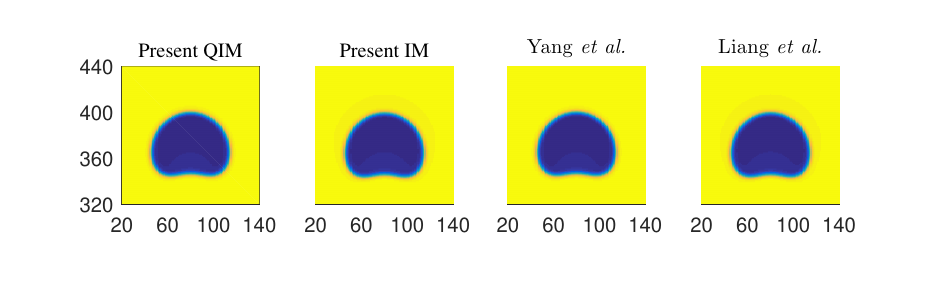}
\caption{(Color online) The enlarged images of density distributions
of the rising bubble at $t=40000$ for present QIM, present IM, model
of Yang \emph{et al.} \cite{Yang}, and model of Liang \emph{et al.}
\cite{Liang}.}
 \label{bubble2}
\end{figure}

\begin{figure}[h]
\centering
\includegraphics[width=0.7\textwidth]{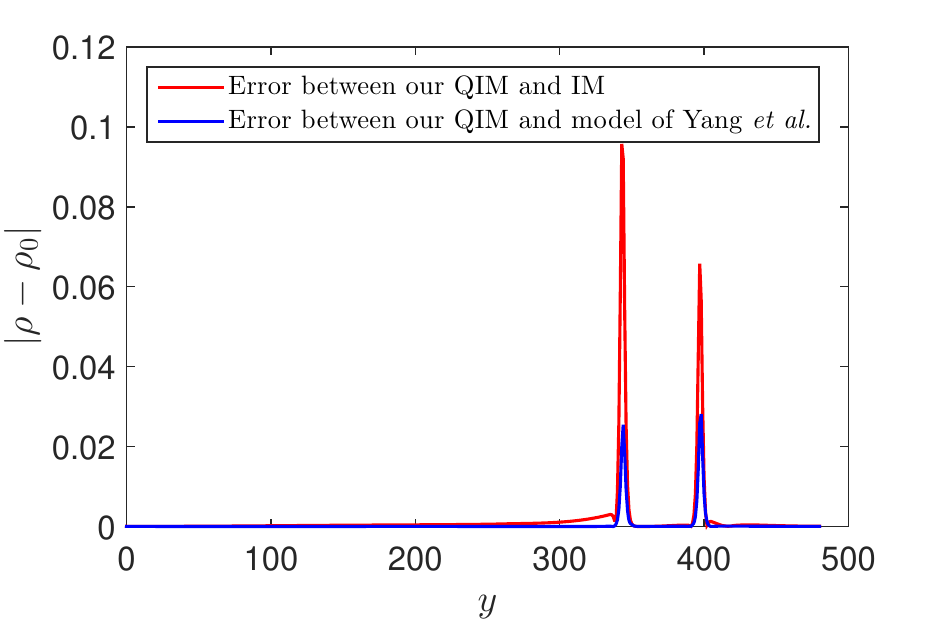}
\caption{(Color online) Errors along the vertical centerline at
$t=40000$,  where $\rho_0$ is the result of our QIM.}
 \label{bubble3}
\end{figure}

To further demonstrate the accuracy of the present models for more
complex flows, the problem of single bubble rising flow driven by
the buoyancy is also considered here. Initially, a circular bubble
(fluid B) without initial velocity is immersed in the bottom center
of another fluid (A). The radius of the initial bubble, $R$,
occupies $32$ lattice spaces. To generate the buoyancy effects, a
body force, $F_{b,y}=-(\rho-\rho_A)g$, is added to the momentum
equation, where $g$ is the gravitational acceleration. We conducted
some simulations on a uniform computational mesh with the size of
$NX\times NY=160\times 480$, and the periodic boundary conditions
are applied at all boundaries. The other related parameters are
given as $W=4, g=10^{-5}, \sigma=0.001, \tau_f=\tau_g=1,
M_{\phi}=6.667, \rho_A/\rho_B=2$. Fig. {\ref{bubble1}} shows the
density distributions of the rising bubble at different times based
on our models and models of Yang \emph{et al.} \cite{Yang} and Liang
\emph{et al.} \cite{Liang}. From this figure, it can be observed
that the results of these models are quite similar to each other.
However, a zoom-in-view of  Fig. {\ref{bubble1}} at $t=40000$ shows
that the incompressible models (our IM and Liang's model) will
produce numerical oscillation near the interface region (see Fig.
{\ref{bubble2}}). To see the differences among these LB models more
clearly, we show the errors along the vertical centerline at
$t=40000$ in Fig. {\ref{bubble3}}, where the error between  our QIM
and IM increases sharply to $0.096$ near the interface. This
phenomenon is reasonable because the compressible term $\gamma
\nabla \cdot (M_{\phi}\nabla \mu)$ has an influence on the numerical
results. While, the maximum error between our QIM and the one in
Ref. \cite{Yang} is approximately $0.028$, which is caused by the
difference between the two models. In the model of Yang \emph{et
al.} \cite{Yang}, the N-S equations in artificial compressible form
can be derived. To neglect the artificial compressible effect, the
condition $T \gg L/c_s$ should be satisfied. While from present QIM,
the quasi-incompressible N-S equations can be exactly recovered in
the limit of a low Mach number.

\begin{figure}[h]
\subfigure[]{ \label{fig:mini:subfig:a7}
\includegraphics[width=2.5in]{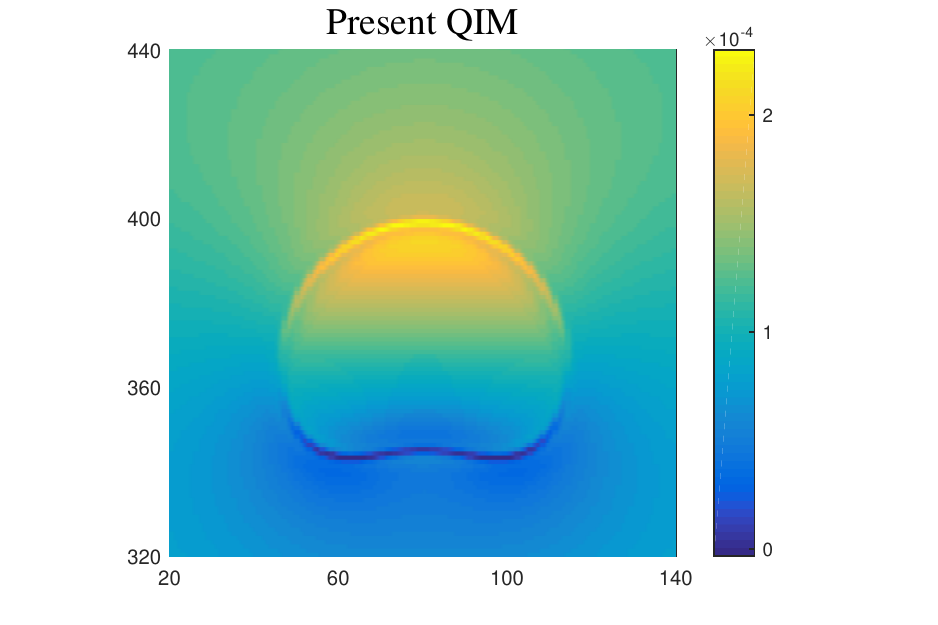}}
\subfigure[]{ \label{fig:mini:subfig:b7}
\includegraphics[width=2.5in]{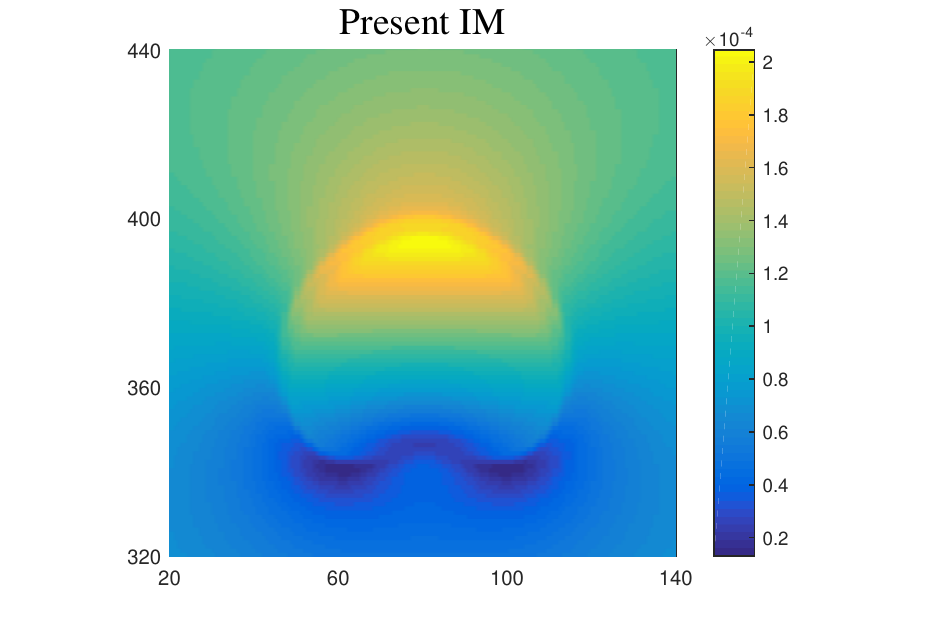}}
\subfigure[]{ \label{fig:mini:subfig:c7}
\includegraphics[width=2.5in]{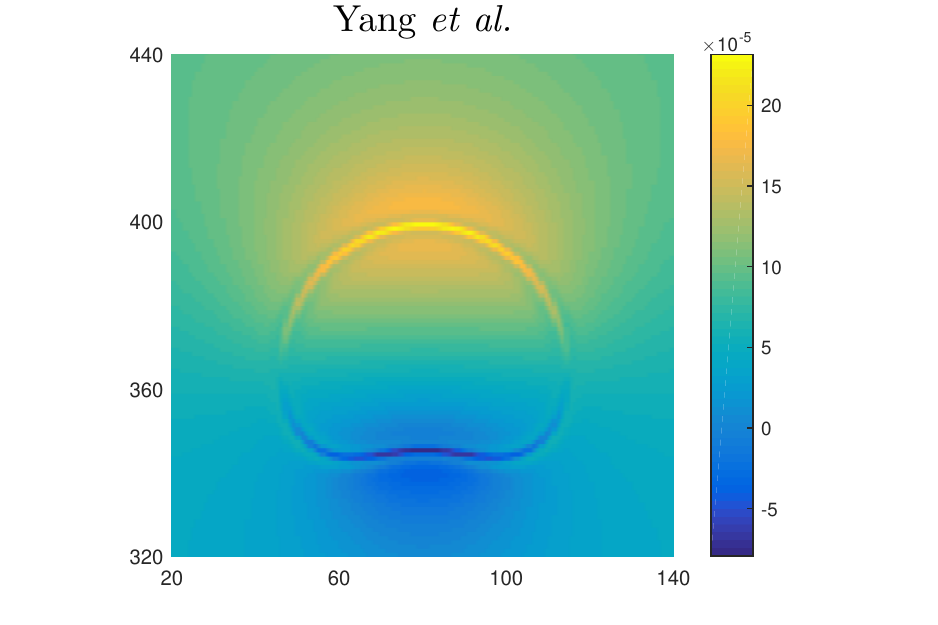}}
\subfigure[]{ \label{fig:mini:subfig:d7}
\includegraphics[width=2.5in]{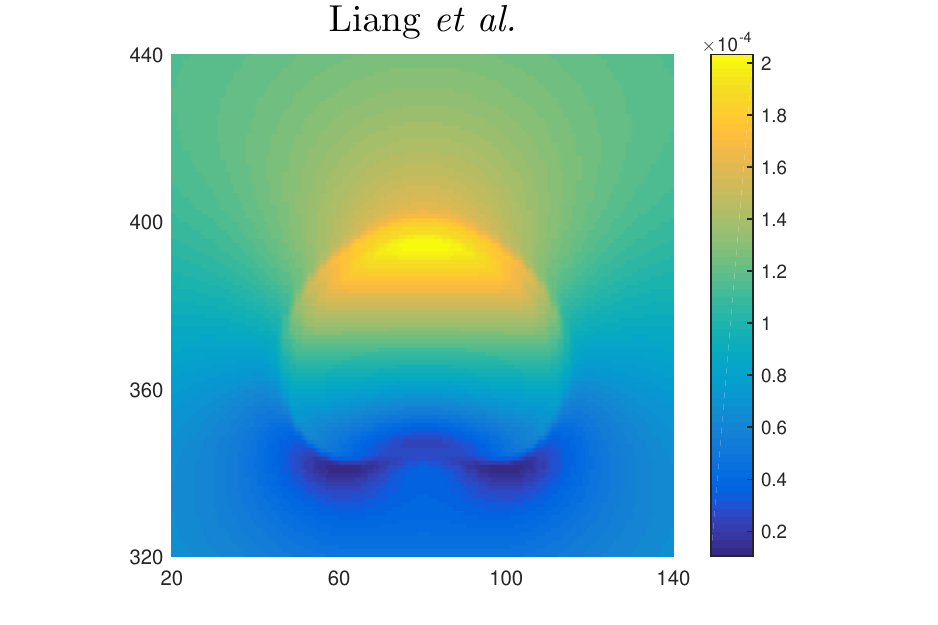}}
\caption{(Color online) Pressure distributions of the rising bubble
at $t=40000$ for (a) present QIM, (b) present IM, (c) model of Yang
\emph{et al.} \cite{Yang}, and (d) model of Liang \emph{et al.}.
\cite{Liang}}
 \label{bubble4}
\end{figure}

Fig. {\ref{bubble4}} depicts the pressure  distributions of the
rising bubble at $t=40000$ based on different models. From this
figure, one can see that there is a significant difference between
quasi-incompressible models and incompressible models, and the
pressure interface of quasi-incompressible models is clearer.

Based on above observations, one can conclude that
quasi-incompressible models (our QIM and Liang's model) are more
superior for complex two-phase flows.

\section{Conclusions}
In this study, to solve single-phase flow problems with a mass
source term in the governing equations and other problems coupled
with the flow field, such as two-phase flow problems, we developed a
generalized LB model for incompressible and nearly incompressible
N-S equations with a mass source term. The proposed model not only
contains some existing models, but also extends them. From the
generalized model, we can not only get some existing models, but
also derive new models. Among these derived models, we can get an
incompressible model for N-S equations with a mass source term, and
we present a modified scheme to calculate the pressure $p$, which is
more accurate than the previous one. Simultaneously, our generalized
model can recover the macroscopic equations without any unnecessary
assumptions. Based on the generalized LB model, a new LB model is
proposed for the quasi-incompressible and incompressible phase-field
system. To validate the accuracy of the proposed model, a series of
numerical tests were performed.

First, we conducted a detailed comparison between present scheme and
the previous one \cite{Liang,Liang2018,Ren}. The result shows that
there is a significant difference between the two pressure schemes
when $S_1 \ne 0$, and theoretically, the present pressure scheme is
more accurate. Then we investigated two basic steady problems of a
static droplet and layered Poiseuille flows. The results of the
former case show that present quasi-incompressible model usually
performs better than the quasi-incompressible model of Yang \emph{et
al.} \cite{Yang} in terms of accuracy, and quasi-incompressible
models are more accurate than incompressible models. In the latter
case, we simulated the layered Poiseuille flows with $\rho_A/\rho_B
=1$ and $\rho_A/\rho_B \ne 1$, and found that our models can obtain
satisfactory results in the velocity under different viscosity
ratios, and our models have a better performance in the interface
region when $\rho_A/\rho_B \ne 1$. Finally, we carried out some
simulations of single bubble rising flow driven by the buoyancy to
further demonstrate the accuracy of the present models. The results
indicate that the incompressible LB models will produce numerical
oscillation near the interface region. While, the
quasi-incompressible LB models seem more reasonable from physical
point of view, and should be considered in the study of complex
two-phase flows.

\section*{Acknowledgements}
The authors are grateful to referees for their valuable comments and
suggestions. This work is supported by the National Natural Science
Foundation of China (Grant Nos. 51576079, 51836003), and the
National Key Research and Development Program of China (Grant No.
2017YFE0100100).

\appendix

\section{\label{app:sec1}Chapman-Enskog analysis of the present model}
 In the Appendix A, we would present the details on how to obtain the proposed LB model for hydrodynamic equations [Eq.
 (\ref{eq:2a})] through the Chapman-Enskog (C-E) expansion.

 Before performing C-E expansion, we first define the zeroth to second moments of the equilibrium distribution
 function,
 \begin{equation}
 \sum_i g_i^{eq}=M_0,\sum_i \mathbf{c}_i g_i^{eq}=M_1, \sum_i \mathbf{c}_i \mathbf{c}_i
 g_i^{eq}=M_2, \sum_i \mathbf{c}_i \mathbf{c}_i \mathbf{c_i}
 g_i^{eq}=M_3.
 \label{eq:a1}
 \end{equation}

In the C-E analysis, the time and space derivatives, the force and
source term can be expanded as,
\begin{subequations}
\begin{equation}
g_i=g_i^{(0)}+\epsilon g_i^{(1)}+\epsilon^{2}g_i^{(2)}+ \cdots,
\label{eq:a2}
\end{equation}
\begin{equation}
G_i=\epsilon G_{i}^{(1)}+\epsilon^2 G_{i}^{(2)}, \label{eq:a3}
\end{equation}
\begin{equation}
\partial_t=\epsilon \partial_{t_1}+\epsilon^2 \partial_{t_2},
\partial_{\alpha}=\epsilon \partial_{1\alpha},
\label{eq:a4}
\end{equation}
\begin{equation}
F_{\alpha}=\epsilon F_{\alpha}^{(1)}+\epsilon^2 F_{\alpha}^{(2)},
\label{eq:a5}
\end{equation}
\begin{equation}
S=\epsilon S^{(1)}+\epsilon^2 S^{(2)},
\end{equation}
\label{eq:A2}
\end{subequations}
 where $\epsilon$ is a small expansion parameter and Greek
indices denote Cartesian spatial components. Using the Taylor
expansion to Eq. (\ref{eq:24}), we have
\begin{equation}
\Delta t D_i g_i(\textbf{x},t)+\frac{\Delta t^2}{2} D_i^2
g_i(\textbf{x},t) + \cdots=-\frac{1}{\tau_g}
\left(g_i(\mathbf{x},t)-g_i^{eq}(\mathbf{x},t)\right)+\Delta
t(1-\frac{1}{2\tau_g})G_i, \label{eq:A2A}
\end{equation}
where $D_{i}=\partial_{t}+c_{i\alpha} \partial_{\alpha}$, and
substituting Eq. (\ref{eq:A2}) into Eq. (\ref{eq:A2A}), one can
obtain the following multi-scale equations,
\begin{subequations}
\begin{equation}
O(\epsilon^0):g_i^{(0)}=g_i^{eq}, \label{eq:a6}
\end{equation}
\begin{equation}
O(\epsilon^1):D_{1i}g_i^{(0)}=-\frac{1}{\tau_g \Delta
t}g_i^{(1)}+(1-\frac{1}{2\tau_g})G_i^{(1)}, \label{eq:a7}
\end{equation}

\begin{equation}
O(\epsilon^2):\partial_{t_2}g_i^{(0)}+D_{1i}g_i^{(1)}+\frac{\Delta
t}{2}D_{1i}^2 g_i^{(0)}=-\frac{1}{\tau_g \Delta t}g_i
^{(2)}+(1-\frac{1}{2\tau_g})G_i^{(2)}, \label{eq:a8}
\end{equation}
\label{eq:A3}
\end{subequations}
where $D_{1i}=\partial_{t_1}+c_{i\alpha} \partial_{1\alpha}$.

Then, the substitution of Eq. (\ref{eq:a7}) into Eq. (\ref{eq:a8})
yields
\begin{equation}
\partial_{t_2}g_i^{(0)}+(1-\frac{1}{2\tau_g})D_{1i}g_i^{(1)}+\frac{\Delta
t}{2}(1-\frac{1}{2\tau_g})D_{1i}G_i^{(1)}=-\frac{1}{\tau_g \Delta
t}g_i ^{(2)}+(1-\frac{1}{2\tau_g})G_i^{(2)}. \label{eq:a9}
\end{equation}
By summing Eq. (\ref{eq:a7}) and Eq. (\ref{eq:a7})$\times
c_{i\beta}$ over $i$, the recovered equations at $\epsilon$ scale
 can be obtained,
\begin{subequations}
\begin{equation}
\partial_{t_1}M_0+\partial_{1\alpha} M_{1\alpha}=-\frac{1}{\tau_g \Delta t}
\sum_i g_i^{(1)}+(1-\frac{1}{2\tau_g})\sum_i G_i^{(1)},
\label{eq:a10}
\end{equation}
\begin{equation}
\partial_{t_1}M_{1\beta}+\partial_{1\alpha} M_{2\alpha \beta}=-\frac{1}{\tau_g \Delta t}
\sum_i c_{i\beta} g_i^{(1)}+(1-\frac{1}{2\tau_g})\sum_i c_{i\beta}
G_i^{(1)}. \label{eq:a11}
\end{equation}
\label{eq:A4}
\end{subequations}

Similarly, we can also obtain the recovered equations  at
$\epsilon^2$ scale from Eq. (\ref{eq:a9})
\begin{subequations}
\begin{equation}
\begin{split}
&\partial_{t_2}M_0+(1-\frac{1}{2\tau_g})\left[\partial_{t_1} (\sum_i
g_i^{(1)})+\partial_{1\alpha}(\sum_i c_{i\alpha}
g_i^{(1)})\right]+\frac{\Delta
t}{2}(1-\frac{1}{2\tau_g})\left[\partial_{t_1} (\sum_i G_i^{(1)})+
\right.
\\ & \left.\partial_{1\alpha}(\sum_i c_{i\alpha} G_i^{(1)})\right]
=-\frac{1}{\tau_g \Delta t}\sum_i
g_i^{(2)}+(1-\frac{1}{2\tau_g})\sum_i G_i^{(2)},
\end{split}
\label{eq:a12}
\end{equation}
\begin{equation}
\begin{split}
&\partial_{t_2}M_{1\beta}+(1-\frac{1}{2\tau_g})\left[\partial_{t_1}
(\sum_i c_{i\beta} g_i^{(1)})+\partial_{1\alpha}\Lambda^{(1)}
\right]+\frac{\Delta t}{2}(1-\frac{1}{2\tau_g})\left[\partial_{t_1}
(\sum_i c_{i\beta}
G_i^{(1)})+\right.\\
& \left. \partial_{1\alpha}(\sum_i c_{i\alpha} c_{i\beta}
G_i^{(1)})\right] =-\frac{1}{\tau_g \Delta t}\sum_i c_{i\beta}
g_i^{(2)}+(1-\frac{1}{2\tau_g})\sum_i c_{i\beta} G_i^{(2)},
\end{split}
\label{eq:a13}
\end{equation}
\label{eq:A5}
\end{subequations}
where $\Lambda^{(1)}=\sum_i c_{i\alpha} c_{i\beta} g_i^{(1)}$ is the
first-order momentum flux tensor.

Summing Eq. (\ref{eq:23a}) and Eq. (\ref{eq:23a})$\times
c_{i\alpha}$ over $i$, one can obtain
\begin{subequations}
\begin{equation}
M_0=\sum_i g_i+\frac{\Delta t}{2}\sum_i G_i, \label{eq:a14}
\end{equation}
\begin{equation}
M_{1\alpha}=\sum_i c_{i\alpha} g_i+\frac{\Delta t}{2} \sum_i
c_{i\alpha} G_i, \label{eq:a15}
\end{equation}
\label{eq:A6}
\end{subequations}
which can be further recast as
\begin{subequations}
\begin{equation}
\sum_i g_i^{(1)} =-\frac{\Delta t}{2} \sum_i G_i^{(1)},\sum_i
g_i^{(2)} =-\frac{\Delta t}{2} \sum_i G_i^{(2)}, \label{eq:a16}
\end{equation}
\begin{equation}
\sum_i c_{i\alpha}g_i^{(1)} =-\frac{\Delta t}{2} \sum_i c_{i\alpha}
G_i^{(1)},\sum_i c_{i\alpha}g_i^{(2)} =-\frac{\Delta t}{2} \sum_i
c_{i\alpha}G_i^{(2)}. \label{eq:a17}
\end{equation}
\label{eq:A7}
\end{subequations}
Substituting Eq. (\ref{eq:A7}) into Eqs. (\ref{eq:A4}) and
(\ref{eq:A5}), we have
\begin{subequations}
\begin{equation}
\partial_{t_1}M_0+\partial_{1\alpha}M_{1\alpha}=\sum_i G_i^{(1)},
\label{eq:a18}
\end{equation}
\begin{equation}
\partial_{t_1}M_{1\beta}+\partial_{1\alpha} M_{2\alpha \beta}=\sum_i c_{i\beta}
G_i^{(1)}, \label{eq:a19}
\end{equation}
\label{eq:A8}
\end{subequations}

\begin{subequations}
\begin{equation}
\partial_{t_2}M_0=\sum_i G_i^{(2)},
\label{eq:a20}
\end{equation}

\begin{equation}
\partial_{t_2}M_{1\beta}+(1-\frac{1}{2\tau_g})\partial_{1\alpha}\Lambda^{(1)} +\frac{\Delta
t}{2}(1-\frac{1}{2\tau_g})\partial_{1\alpha}(\sum_i c_{i\alpha}
c_{i\beta} G_i^{(1)}) =\sum_i c_{i\beta} G_i^{(2)}. \label{eq:a21}
\end{equation}
\label{eq:A9}
\end{subequations}
Combining Eq. (\ref{eq:a18}) and Eq. (\ref{eq:a20}) at $\epsilon$
and $\epsilon^2$ scales yields
\begin{equation}
\partial_t M_0+\partial_{1\alpha}M_{1\alpha}=\sum_i G_i.
\label{eq:a22}
\end{equation}

To recover the continuity equation with a source term [Eq.
(\ref{eq:1})], the following conditions should be satisfied
\begin{equation}
M_0=\tilde {\rho},M_1=\rho \mathbf{u},\sum_i G_i=S. \label{eq:a23}
\end{equation}
In addition, to recover the momentum equation [Eq. (\ref{eq:2})],
$M_2=p\mathbf{I}+\rho \mathbf{uu}$ is also needed. Thus, the
equilibrium distribution function can be given as
\begin{equation}
g_i^{eq}=\left\{
\begin{array}{l}
\tilde{\rho}+\frac{p}{c_s^2}(\omega_i-1)+\rho s_i(\mathbf{u}), \;\;\;i=0,\\
\frac{p}{c_s^2}\omega_i+\rho s_i(\mathbf{u}),
\;\;\;\;\;\;\;\;\;\;\;\;\;\;\;\;\;   i \ne 0,
\end{array}
\right. \label{eq:a23a}
\end{equation}
and Eq. (\ref{eq:a19}) can be rewritten as
\begin{equation}
\partial_{t_1} (\rho
u_{\beta})+\partial_{1\beta}p+\partial_{1\alpha}(\rho
u_{\alpha}u_{\beta})=\sum_i c_{i\beta}G_i^{(1)}. \label{eq:a19a}
\end{equation}
To derive the equation at $\epsilon^2$ scale, we first express
$\Lambda^{(1)}$  as
\begin{equation}
\begin{split}
\Lambda^{(1)}&=-\tau_g \Delta t \left[ \partial_{t_1}M_{2\alpha
\beta}+\partial_{1\gamma}M_{3\alpha \beta
\gamma}-(1-\frac{1}{2\tau_g})\sum_i c_{i\alpha} c_{i\beta}
G_i^{(1)} \right] \\
&=-\tau_g \Delta t \left \{\partial_{t_1}
p\delta_{\alpha\beta}+\partial_{t_1}(\rho
u_{\alpha}u_{\beta})+c_s^2\partial_{1\gamma}( \rho
u_{\gamma}\delta_{\alpha\beta})+c_s^2
\partial_{1\gamma}[\rho(u_{\alpha \delta_{\beta \gamma}}+u_{\beta}\delta_{\alpha
\gamma})] \right.\\
&\left.\quad -(1-\frac{1}{2\tau_g})\sum_i
c_{i\alpha}c_{i\beta}G_i^{(1)}\right \},
\end{split}
\label{eq:a24}
\end{equation}
where Eq. (\ref{eq:a7}) has been used, and the term
$\partial_{t_1}(\rho u_{\alpha} u_{\beta})$ can be given by
\begin{equation}
\partial_{t_1}(\rho u_{\alpha}
u_{\beta})=u_{\alpha}(\sum_ic_{i\beta}G_i^{(1)})+u_{\beta}(\sum_ic_{i\alpha}G_i^{(1)})-(u_{\alpha}
\partial_{1\beta}p+u_{\beta}
\partial_{1\alpha}p)-u_{\alpha}u_{\beta}\tilde S^{(1)},
\label{eq:a25}
\end{equation}
where the Eqs. (\ref{eq:a19a}) and Eq. (\ref{eq:a18}) are used,
$\tilde S^{(1)}=S^{(1)}+\partial_{t_1} (\rho-\tilde{\rho})$ and the
term of $O(Ma^3)$ has been neglected.

Combining Eq. (\ref{eq:a19a}) with Eq. (\ref{eq:a21}) at $\epsilon$
and $\epsilon^2$ scales, together  with Eq. (\ref{eq:a24}) and Eq.
(\ref{eq:a25}), we now obtain
\begin{equation}
\begin{split}
\partial_t (\rho u_{\beta})+\partial_{\alpha}(\rho
u_{\alpha}u_{\beta})=&-\partial_{\beta}p+\partial_{\alpha}\left[
\rho
\nu(\partial_{\alpha}u_{\beta}+\partial_{\beta}u_{\alpha})\right]+\epsilon
\Delta t (\tau_g-0.5)\partial_{\alpha} \left [ \partial_{t_1}p
\delta_{\alpha \beta}\right.
 \\ &+\partial_{1\gamma}(c_s^2\rho
u_{\gamma}\delta_{\alpha
\beta})+u_{\alpha}(\sum_ic_{i\beta}G_i^{(1)})+u_{\beta}(\sum_ic_{i\alpha}G_i^{(1)})+\\
&
c_s^2(u_{\alpha}\partial_{1\beta}\rho+u_{\beta}\partial_{1\alpha}\rho)-(u_{\alpha}
\partial_{1\beta}p+ u_{\beta}
\partial_{1\alpha}p)-u_{\alpha}u_{\beta}\tilde{S}^{(1)}\\ & \left.-\sum_i
c_{i\alpha}c_{i\beta}G_i^{(1)} \right ] +\sum_i c_{i\beta}G_i.
\end{split}
\end{equation}
where the kinetic viscosity is determined by
\begin{equation}
\nu=c_s^2(\tau_g-0.5)\Delta t.
\end{equation}

Compared to Eq. (\ref{eq:2}), we can recover the momentum
 equation as long as the following equations hold,
\begin{equation}
\sum_i c_{i\beta}G_i=F_{\beta}, \label{eq:a25aa}
\end{equation}
\begin{equation}
\begin{split}
\sum_i c_{i\alpha}c_{i\beta}G_i=&\partial_{t}p \delta_{\alpha
\beta}+\partial_{\gamma}(c_s^2\rho u_{\gamma}\delta_{\alpha
\beta})+u_{\alpha}F_{\beta}+u_{\beta}F_{\alpha}+c_s^2(u_{\alpha}\partial_{\beta}\rho+u_{\beta}\partial_{\alpha}\rho)-\\&(u_{\alpha}
\partial_{\beta}p+ u_{\beta}
\partial_{\alpha}p)-u_{\alpha}u_{\beta}\tilde{S}+\left(\frac{2}{d}\rho
c_s^2-\frac{\xi}{\Delta t(\tau_g-0.5)}\right)
\partial_{\gamma}u_{\gamma}\delta_{\alpha \beta}. \label{eq:a25aaa}
\end{split}
\end{equation}

Based on Eqs. (\ref{eq:a25aa}), (\ref{eq:a25aaa}), and the last
equation in Eq. (\ref{eq:a23}),  the force distribution function can
be given by
\begin{equation}
G_i=\omega_i\left\{ S+\frac{\mathbf{c}_i \cdot
\mathbf{F}}{c_s^2}+\frac{(\mathbf{c}_i
\mathbf{c}_i-c_s^2\mathbf{I}):\left[\partial_t(p-\tilde{\rho}c_s^2)\mathbf{I}+\mathbf{u
\tilde{F}}+\mathbf{ \tilde{F}u}-\mathbf{uu}\tilde{S}+Q (\nabla \cdot
\mathbf{uI})\right]}{2c_s^4}\right \} , \label{eq:a27}
\end{equation}
where $\tilde{F}$ is a modified total force
\begin{equation}
\mathbf{\tilde{F}}=\mathbf{F}-\nabla(p-\rho c_s^2),
\end{equation}
$\tilde{S}$ and $Q$ can be expressed as
\begin{equation}
\tilde{S}=S+\partial_t(\rho-\tilde{\rho}).
\end{equation}
\begin{equation}
Q=\frac{2}{d}{\rho} c_s^2-\frac{\xi}{\Delta t(\tau_g-0.5)}.
\end{equation}

\section{\label{app:sec2}The computation of the pressure}
Now we will focus on how to calculate the pressure from the
distribution function $g_i$. According to the expression of
$g_0^{eq}$, we have
\begin{equation}
\frac{p}{c_s^2}(1-\omega_0)=\tilde{\rho}+\rho
s_0(\mathbf{u})-g_0^{eq}. \label{eq:b0}
\end{equation}
Actually, once $g_0^{eq}$ in Eq. (\ref{eq:b0}) is replaced by the
distribution function $g_i$, one can present a scheme to calculate
pressure. Firstly, from Eq. (\ref{eq:a7}) we can derive
\begin{equation}
\epsilon g_i^{(1)}=-\epsilon \tau_g \Delta t \left[
D_{1i}g_i^{(0)}-(1-\frac{1}{2\tau_g})G_i^{(1)}\right]+O(\Delta t^2),
\end{equation}
or equivalently,
\begin{equation}
g_i-g_i^{eq}=-\epsilon \tau_g \Delta t \left[
D_{1i}g_i^{(0)}-(1-\frac{1}{2\tau_g})G_i^{(1)}\right]+O(\Delta t^2).
\label{eq:b1}
\end{equation}
Taking the zeroth-direction of Eq. (\ref{eq:b1}), we have
\begin{equation}
\begin{split}
g_0-g_0^{eq}&=-\epsilon \tau_g \Delta t \left[
\partial_{t_1}g_0^{(0)}-(1-\frac{1}{2\tau_g})G_0^{(1)}\right]+O(\Delta t^2)\\ &= - \tau_g \Delta t \left[
\partial_{t}g_0^{(0)}-(1-\frac{1}{2\tau_g})G_0\right]+O(\Delta t^2).
\end{split}
\end{equation}
Note that the term $\partial_t g_0^{(0)}$ is the order of $O(Ma^2)$,
thus, we get
\begin{equation}
-g_0^{eq}=-g_0+\tau_g \Delta t(1-\frac{1}{2\tau_g})G_0+O(\Delta
t^2+\Delta t Ma^2). \label{eq:b2}
\end{equation}
Neglecting the terms of $O(\Delta t^2+\Delta t Ma^2)$, and
substituting Eq. (\ref{eq:b2}) into Eq. (\ref{eq:b0}), we can obtain
\begin{equation}
\begin{split}
\frac{p}{c_s^2}(1-\omega_0)&=\tilde{\rho}+\rho
s_0(\mathbf{u})-g_0+\Delta
t(\tau-\frac{1}{2})G_0\\&=\tilde{\rho}+\rho s_0(\mathbf{u})-(\sum_i
g_i-\sum_{i \ne 0} g_i)+\Delta
t(\tau-\frac{1}{2})G_0\\&=\tilde{\rho}+\rho s_0(\mathbf{u})-(\sum_i
g_i^{eq}-\frac{\Delta t}{2}\sum_i G_i-\sum_{i \ne 0} g_i)+\Delta
t(\tau-\frac{1}{2})G_0\\&=\tilde{\rho}+\rho
s_0(\mathbf{u})-\tilde{\rho}+\frac{\Delta t}{2}S+\sum_{i \ne 0}
g_i+\Delta t(\tau-\frac{1}{2})G_0
\\&=\rho
s_0(\mathbf{u})+\frac{\Delta t}{2}S+\sum_{i \ne 0} g_i+\Delta
t(\tau-\frac{1}{2})G_0.
\end{split}
\end{equation}
As a result, the pressure can be calculated as
\begin{equation}
p=\frac{c_s^2}{1-\omega_0} \left[ \sum_{i\ne 0} g_i +\frac{\Delta
t}{2}S+\rho s_0(\mathbf{u})+\Delta t(\tau-\frac{1}{2})G_0 \right ],
\label{eq:b3}
\end{equation}
which has an accuracy of $O(\Delta t^2+\Delta t Ma^2)$.


\end{document}